# Intermediates of Forming Transition Metal Dichalcogenide Heterostructures Revealed by Machine Learning Simulations

Luneng Zhao[1,2], Hongsheng Liu[1], Yuan Chang[1], Xiaoran Shi[1], Jijun Zhao[3], Feng Ding[2], Junfeng Gao[1,2*]

[1] Key Laboratory of Materials Modification by Laser, Ion and Electron Beams (Dalian University of Technology), Ministry of Education, School of Physics, Dalian 116024, China

[2] Suzhou Laboratory, Suzhou 215123, China

[3] Guangdong Basic Research Center of Excellence for Structure and Fundamental Interactions of Matter, Guangdong Provincial Key Laboratory of Quantum Engineering and Quantum Materials, School of Physics, South China Normal University, Guangzhou 510006, China

Correspondence to: Junfeng Gao, gaojf@dlut.edu.cn

## Abstract

Two-dimensional (2D) transition metal dichalcogenide (TMD) van der Waals heterostructures (vdWHs) hold promise for high-performance electronics, but their large-scale synthesis remains limited by size constraints and alloying contaminations. Recently, a two-step vapor deposition method was reported for growing wafer-size TMD vdWHs with minimal impurities. In this study, we develop a machine learning potential (MLP) that accurately captures the atomic-scale dynamic growth process of bilayer $MoS_2/WS_2$ vdWHs under feasible growth conditions. Our simulations uncover a crucial metastable SMMS (M = Mo or W) intermediate structure that facilitates metal atom swap and alloying. Eliminating the alloying contamination requires preventing the embedding of bare metal atoms. The results also show that the SMMS structure exhibits favourable electronic properties and emerges as a low Schottky barrier contact electrode for $MoS_2$ field-effect transistors (FETs).

## Introduction

2D TMDs have been extensively studied due to their semiconducting band gaps and high carrier mobility [1–4], strong nonlinear optical response [5,6], and ease of layer stacking assembly. The integration of vdWHs by pristine TMD layers can further enhance their properties, enabling various potential applications in microelectronics, optoelectronics, and nonlinear optics [7–14].

Despite their promise, the controlled growth of TMD vdWHs faces numerous challenges. While commonly used mechanical assembly methods can achieve high-quality TMD vdWHs with atomically sharp interfaces [15,16], they often struggle to achieve wafer-size dimensions and can be prohibitively expensive. In contrast, chemical vapor deposition (CVD) has achieved significant success in growing wafer-sized monolayer TMD [12,17,18]. However, growing TMD vdWHs still faces limitations in size constraints and a strong tendency toward alloy formation [19,20]. Among these scalable methods, metal-organic chemical vapor deposition (MOCVD) has emerged as a promising alternative for growing large-size TMD heterojunctions with improved interfaces [20–22].

Recently, it was reported that a two-step vapor deposition process with a high-to-low temperature strategy was used to synthesize wafer-sized TMD vdWHs [23]. In this method, a monolayer $WS_2$ film was first achieved by depositing a W film on a sapphire substrate and sulfurizing it at 900 °C, the highest temperature among the four stacked materials. Next, a Mo film was deposited on the $WS_2$ monolayer via magnetron sputtering and sulfurized at 800 °C to form $MoS_2$. Subsequently, a Nb film was deposited on the $WS_2/MoS_2$ film via magnetron sputtering and selenized at 700 °C to form $NbSe_2$. Finally, $PtTe_2$ was grown on the $WS_2/MoS_2/NbSe_2$ vdWHs at 350 °C, the lowest temperature in the sequence. The final structure was a wafer-sized vdWHs several centimetres in size, consisting of four layers: $WS_2/MoS_2/NbSe_2/PtTe_2$. The authors further proposed that, during the metal deposition process, the metal atom films coat the TMD surface[23].

Compared with experimental trials that involve many parameters and various growth conditions, accurate atomic simulations can certainly provide an in-depth understanding of the growth of TMD vdWHs. The growth of TMD vdWHs is always accompanied by rapid formation and breaking, involving various intermediate motifs with complex chemical bonds, such as metallic bonds, covalent bonds, both covalent and ionic characters of Mo-S bonds, and layered vdW forces. Density functional theory (DFT) is well suitable for describing the complex chemical bonding, but is limited to afford large-scale simulations. In contrast, classical molecular dynamics (MD) can simulate large systems but lacks the capability to capture complex chemical bond recombination. Although various machine learning potentials (MLPs) have been developed and reported to simulate large systems with accuracy comparable to DFT [24–29], only a few MLPs are capable of describing intricate growth processes of TMD vdWHs.

In this study, we developed an MLP trained on extensive DFT data and utilizing a revised equivariant graph neural network implemented in the NequIP package [24]. The MLP was then implemented into MD (MLP-MD), enabling both large-scale and accurate simulation of the $MoS_2/WS_2$ vdWHs growth process. Our results indicate that a bare metal atomic layer is unstable on TMD layers; it spontaneously sinks into the S layer, forming a crucial intermediate structure (SMMS) with high stability. This behaviour facilitates the exchange between Mo and W atoms, thereby revealing the atomic-scale mechanism of TMD alloys formation. To grow pristine TMD vdWHs, it is essential to suppress the SMMS formation, which can be achieved by preventing bare metal atom adsorption on pre-existing TMD layers. On the other hand, the SMMS structure serves as an ideal metallic electrode with a low Schottky barrier height, enabling $MoS_2$ integrated circuits through the planar deposition of metal atoms on uncovered $MoS_2$.

# Results

## MLP Development and DFT Validation

To enhance the complexity and versatility of our MLP model, we designed a diverse dataset encompassing a wide range of Mo, W, S, and mixed structures. We also included Se atomic structures to broaden the scope of potential research applications. Through an iterative learning process, we carefully selected and balanced the weights of different structural configurations. The initial training set was derived from on-the-fly machine learning MD simulations[30] and was then continuously enriched with new structures throughout successive iterations. The final dataset included various TMD layers, $MoS_2/WS_2$ vdWHs, $MoS_2$ or $WS_2$ with homogeneous and heterogeneous metal clusters, S clusters, diverse MoWS alloy configurations, and growth intermediate structures [Fig. 1]. This comprehensive approach resulted in a dataset of approximately 26,000 DFT-calculated entries. Detailed procedures for both the initial training and subsequent iterative refinement are provided in the Supplementary Information. Figure S1(a) demonstrates a high consistency between the MLP and DFT energies, achieving a root mean square error (RMSE) of 10.6 meV/atom. Figure S1(b) further illustrates a high consistency between atomic forces predicted by the MLP and those from DFT calculations, with an RMSE of 151 meV/Å. To comprehensively assess the model's reliability across the energy landscape, we analysed the prediction errors as a function of structure formation energy [see Fig. S1(c)]. This analysis confirms that the MLP maintains high accuracy not only for stable low-energy configurations but also for high-energy states relevant to growth simulations. Overall, despite the structural complexity and complex bond changes in the growth process potentially introducing some noise, these errors are significantly smaller than the thermal energy ($kT$) of ~90 meV at typical $MoS_2$ growth temperatures (~800 °C). Therefore, our MLP model maintains sufficient accuracy for describing TMD growth processes. These error values are comparable to those reported in recently published machine learning potentials for growth simulations [31,32]. A systematic comparison of energies of key structures obtained from MLP and DFT further confirms the reliability of our model shown in Table S1. It can be seen that a high consistency between MLP and DFT in the relative stabilities of key configurations such as SMMS and $MS_2$. Specifically, for the SMMS-like configuration generated by direct MD deposition and the constructed and optimized SMMS structure, the energy difference is 0.080 eV and 0.068 eV, obtained from DFT calculations and MLP predictions, respectively. Similarly, for the SMoWS system, the corresponding energy difference is 0.090 eV from DFT simulation and 0.077 eV from MLP prediction. These results demonstrate that our MLP can accurately compute the energy differences between different structures and reliably describe the relative stabilities of materials.

To further validate the capability of our MLP to simulate the growth of TMD materials using MLP-MD, we randomly mixed Mo, W, and S atoms in a 1:1:4 ratio, then annealed the mixture at temperatures ranging from 1500 K to 900 K for 2 ns. This process resulted in the formation of ordered TMD layers, including the 1H and 1T phases, demonstrating the reliability of our MLP in simulating the complex layered growth behaviour of TMDs [Fig. S2]. An animation of the annealing process is available in the Supplementary Information as Supplementary Movie 1. These results underscore the effectiveness of our iterative learning approach in developing a robust and accurate MLP for simulating the growth of TMD heterostructures.

## Deposition Dynamics of Mo Atoms on $MoS_2$

Before investigating TMD vdWHs, the developed MLP-MD model successfully simulated the growth of bilayer $MoS_2$ through a two-step vapor deposition method [23]. First, the adsorbed single Mo atom was unstable on the $MoS_2$ layer [Fig. 2(a)] and, at the typical growth temperature of 1100 K, quickly sank beneath the S atom layer within tens of picoseconds [Fig. 2(b)]. DFT calculations confirm that this embedding process releases 1.45 eV of energy. Figure 2(c) shows a significant drop in energy at around 50 ps, indicating that the Mo atom has been embedded into the $MoS_2$ layer. Previous studies have shown that single or paired metal atoms suspended on bare surfaces exhibited significant catalytic enhancement [33–36]. This indicates that the configuration of metal atoms needs to be carefully considered in single-atom catalysis.

Subsequently, individual Mo atoms were simulated to be continuously sputtered onto the existing $MoS_2$ layer with a kinetic energy of 0.12 eV. To simulate the stability during the metal atom deposition and pre-heating stage, simulations were performed at 900 K. The deposition process of Mo atoms that resulted in 0.25 monolayers (MLs) [Fig. 2(d)] and 1.0 ML [Fig. 2(e)] demonstrated the behaviour of Mo atoms on the $MoS_2$ layer [see Supplementary Movie 2]. Figure S3 depicts the evolution of Mo-Mo bonds during continuous deposition, indirectly reflecting the embedding process. We observed that the number of Mo-Mo bonds increased gradually with continued deposition and reached saturation once deposition stopped. As shown in Fig. S4, when Mo atoms were deposited at higher kinetic energies, such as 1 eV and 5 eV, the Mo atoms were still embedded in $MoS_2$. This indicates that the embedding behaviour of Mo atoms is not sensitive to kinetic energy. Throughout the MLP-MD simulation, no Mo atoms formed the MoSMoS structure [Fig. 2(f)]. Instead, all deposited Mo atoms spontaneously sank into the $MoS_2$ layer. These findings suggest that under practical deposition conditions, Mo atoms preferentially form an SMoMoS structure [Fig. 2(g)]. This SMoMoS structure was then constructed and optimized using DFT, revealing that it exhibits an energy 0.080 eV/atom lower than that of the directly formed embedded structure from MD simulations [Fig. 2(e)].

We conducted a detailed comparison of these two structures in terms of energy and dynamic stability. DFT calculations reveal that the energy decreases significantly by 1.73 eV per Mo atom when transitioning from MoSMoS to SMoMoS, indicating a strong thermodynamic driving force towards the formation of the SMoMoS structure. In addition to the SMoMoS structure formed by depositing 1 ML of Mo atoms, we constructed and optimized structures such as $SMo_3S$ [Fig. 2(h)] and $SMo_4S$ [Fig. 2(i)], which may form with an increased number of Mo atoms. The convex hull of MoS compounds was plotted by varying the Mo ratio, referencing the bulk phases of elements Mo and S, as shown in Fig. 2(j). As shown in the convex hull, $MoS_2$ has the lowest energy, while SMoMoS is situated 196 meV/atom above the convex hull. We also found structures that match this elemental ratio from other 2D materials databases (MatHub-2d)[37] [Fig. S5], with their formation energies indicated by green dots in the figure, showing that they are 253 meV/atom higher than the SMoMoS structure. $SMo_3S$ and $SMo_4S$ are 221 meV/atom and 218 meV/atom above the convex hull, respectively, indicating that when approximately 1 ML of Mo atoms is deposited on the $MoS_2$ surface, SMoMoS is the most likely intermediate. The phonon dispersion of the SMoMoS structure showed no imaginary frequencies [Fig. 2(k) and Fig. 2(l)], further confirming that MoSMoS would spontaneously transform into SMoMoS. It is worth noting that the phonon dispersion simulated by the MLP showed good agreement with the DFT results, validating the accuracy of our MLP in describing structural stability. Long-term MLP-MD simulations further validated the stability of SMoMoS, as shown in Fig. S6.

Furthermore, we used DFT simulations to validate the key processes described above. Figure S7(a) illustrates the process of a single Mo atom embedding into the $MoS_2$ layer. The embedding time for vacuum-deposited Mo atoms is approximately 700 fs, contrasting with the ~50 ps required for surface-adsorbed atoms, as indicated by the energy evolution in Fig. S7(a). Figure S7(b) expands our investigation of the behaviour of the Mo atomic layer on $MoS_2$. As the Mo atoms were gradually embedded into the $MoS_2$ structure, they descend accordingly. This energy decrease is significantly lower than that of the initial configuration, further confirming the thermodynamic favourability of Mo atom embedding. Figure S7(c) focuses on the stability of the formed SMoMoS structure over approximately 10 ps.

## Heterogeneous Deposition Dynamics of Mo Atoms on $WS_2$

Following the two-step vapor-deposition process [23], we continuously deposited Mo atoms onto the $WS_2$ monolayer in MLP-MD simulations, leading to the formation of $MoS_2/WS_2$ vdWHs [see Supplementary Movie 3]. Figures 3(a-c) display snapshots from the simulation at 0.06 ns, 0.15 ns, and 1.1 ns, showing that the Mo atoms do not remain on the $WS_2$ surface but are embedded within the $WS_2$ monolayer. Figure S8 shows the evolution of Mo/W-Mo/W bonds during continuous Mo deposition, indirectly reflecting the degree of Mo atom insertion. The number of these bonds gradually increases with deposition and saturates once deposition stops. Importantly, during the simulation, Mo and W atoms in the SMoWS intermediate structure layer are able to exchange [Fig. 3(d-f)], resulting in alloying. The lattice constants of $MoS_2$ and $WS_2$ are nearly identical (approximately 1% [23]), reflecting similar behaviour for Mo and W. At 300 K [see Fig. S10], simulation results show that while Mo atoms continue to embed into the $WS_2$ monolayer to form the SMoWS structure, no Mo-W atomic exchange occurs over a 4.5 ns trajectory, indicating that alloying is kinetically suppressed under low-temperature conditions.

Insight into this sinking and subsequent alloying transformation is obtained by analysing relevant energy and phonon dispersions. A single layer of Mo on $WS_2$ [Fig. 3(g)] is energetically unfavourable compared to the embedded configuration. Due to the similar interaction between Mo atoms and $WS_2$, as previously observed with $MoS_2$, the constructed SMoWS intermediate structure shown in Fig. 3(h) emerges as another crucial configuration to consider. Its energy is 0.090 eV/atom lower than that of the directly formed structure from MD simulations [Fig. 3(c)]. Figure 3(j) shows the presence of imaginary frequencies in the phonon spectrum, indicating that this structure is unstable. Conversely, the Mo atoms completely sink and embed under the top S layer of the $WS_2$ monolayer, forming the SMoWS intermediate structure [Fig. 3(h)], releasing 2.08 eV per Mo atom (based on DFT calculations). In addition to the sinking of Mo atoms, atomic exchange between Mo and W atoms occurs, transforming the SMoWS intermediate structure into an alloyed SMMS structure [Fig. 3(i)]. The phonon spectra of both SMoWS [Fig. 3(k)] and SMMS [Fig. 3(l)] show no imaginary frequencies throughout the Brillouin zone, confirming their dynamical stability. Long-term MLP-MD simulations further validate the stability of the SMMS structure, as shown in Fig. S9(a) and Fig. S9(b).

To determine the thermodynamic driving force for alloying, we calculated the free energy change $\Delta F$ for varying distributions of metal atoms between the upper and lower layers. Here, $\Delta F$ is defined as

$$\Delta F = \frac{1}{N}[E - E_{\mathrm{SMoWS}} - T\Delta S] \tag{1}$$

, where $E_{\mathrm{SMoWS}}$ denotes the energy of the unalloyed SMMS structure [Fig. 3(h)], $E$ is the energy of the alloyed configuration, $N$ is the number of atoms, $T$ is the temperature (300 K), and $\Delta S$ represents the mixing entropy for Mo/W alloying. The mixing entropy $S$ is given by

$$S = (N_{\mathrm{Mo}} + N_{\mathrm{W}})k_b[-x\ln(x) - (1-x)\ln(1-x)] \tag{2}$$

, where $k_b$ is the Boltzmann constant, $x$ is the fraction of Mo atoms in the upper metal layer, and $N_{\mathrm{Mo}}$ and $N_{\mathrm{W}}$ are the numbers of Mo and W atoms, respectively. By randomly exchanging metal atoms between the upper and lower layers in the $(7 \times 7)$ SMMS structure supercell and optimizing the structure, we obtained $\Delta F$ for the SMMS structures with different alloy proportions [Fig. 3(m)]. We found that when the different metal atoms are uniformly distributed between the upper and lower layers, the free energy is minimized; specifically, the energy of S($Mo_{0.5}W_{0.5}$)($W_{0.5}Mo_{0.5}$)S (alloyed SMMS) is approximately 11.3 meV/atom lower than that of the unalloyed SMoWS structure. This energy reduction is primarily attributed to the configurational entropy, contributing 9.0 meV/atom, and the alloy structure also alleviates stress caused by structural asymmetry. Furthermore, the relaxed SMMS structure exhibits a lattice constant approximately 5% smaller than that of monolayer $WS_2$. As a result, the deposition and embedding of sufficient Mo atoms into the $WS_2$ substrate generate accumulated stress, leading to potential cracking of the underlying metal-S layer and subsequently accelerating atomic exchange between Mo and W.

In the above-mentioned study[23], a two-step vapor deposition process was reported to synthesize wafer-scale TMD vdWHs, where the key intermediate structure during growth is a Mo atomic monolayer on the $WS_2$ surface. However, our MLP-MD simulations reveal that the Mo atomic monolayer on the $WS_2$ surface is highly unstable, it can be easily transformed into the SMoWS structure by thermal annealing. This suggests that the experimentally observed clean interfaces must be governed by kinetics or environmental factors that prevent the formation of this bare metal intermediate.

## Sulfurization of Intermediate Structures

We now pose the question of what happens when S is deposited on the SMoMoS and alloyed SMMS intermediate phases during the second step. For SMoMoS, a sufficient amount of S atoms was further deposited on the top surface of the SMoMoS intermediate phase [Fig. S11(a-e) and Supplementary Movie 4]. The simulation shows that the SMoMoS intermediate phase was initially very stable. After 531 ps [Fig. S11(b)], we observed that S atoms penetrated the SMoMoS structure and pulled some Mo atoms to the surface, gradually forming a bilayer of $MoS_2$ [Fig. S11(e)]. As shown in Fig. S12(a), there are very few initial Mo-S bonds because the S atoms deposited on the SMMS surface cannot bond with Mo. As the MD simulation progresses, some Mo atoms are pulled to the surface to bond with the deposited S atoms, leading to an increase in the number of Mo-S bonds. Between 700 and 1000 ps, a large number of Mo atoms are pulled to the surface, resulting in a substantial increase in Mo-S bonds. The number of Mo-S bonds tends to be saturated after about 1.2 ns, when a bilayer $MoS_2$ is fully formed.

Similarly, for the alloyed SMMS intermediate phase, a sufficient amount of S atoms was further deposited on the top surface [see Supplementary Movie 5]. As shown in Fig. 4(a), the alloyed SMMS intermediate phase was also very stable initially. After 608 ps of simulation, S atoms penetrated the SMMS structure and pulled metal atoms from the top surface [Fig. 4(b)]. However, in the alloyed intermediate phase, S atoms did not selectively extract either Mo or W atoms, instead, they pulled both Mo and W atoms from the upper layer to the surface [Fig. 4(c-e)]. Therefore, the resulting structure was not a distinct $MoS_2/WS_2$ vdWHs, but rather an alloyed $Mo_xW_{1-x}S_2/Mo_{1-x}W_xS_2$ vdWHs. As shown in Fig. S12(b), the evolution of Mo/W-S bonds clearly illustrates this process. According to the experimental results of Zhou et al.[23], the successful synthesis of non-alloyed $MoS_2/WS_2$ heterostructures suggests that the formation of the SMMS intermediate state may have been avoided through process control during the experiment. In the scenario where the intermediate state is the SMoWS structure, we further conducted sulfurization MLP-MD simulations. As shown in Fig. S11(f-j), during the simulation, the upper Mo atoms detached from the SMoWS structure. S atoms quickly occupied the original positions of the Mo atoms, effectively preventing the lower W atoms from being extracted. Following the sulfurization of SMoWS, a non-alloyed $MoS_2$ layer formed on the upper layer. This mechanism provides an atomic-level explanation for the experimentally observed non-alloyed growth.

To achieve high-quality, non-alloyed TMD vdWHs, it is essential to prevent Mo atom sinking and the formation of SMMS intermediate phases during growth. We employed MLP-MD to study the behaviour of a single Mo atom and various Mo-S clusters on a $MoS_2$ substrate. A bare Mo atom can quickly sink into the $MoS_2$ monolayer, and once submerged, the Mo atom remains firmly embedded with no surface diffusion observed throughout the simulation [Fig. 4(f-i)]. The trajectory projection in the $xy$ plane [Fig. 4(j)] further confirms this, showing that the Mo atom is trapped at its initial position throughout the 1.4 ns simulation.

In contrast, once a Mo atom bonds with an S atom, the $Mo-S_1$ structure remains on the surface without sinking for the duration of 1.4 ns [Fig. 4(k-n)]. The structure of $Mo-S_1$ exhibits slightly higher surface diffusion (moving one step every 200 ps), primarily between adjacent Mo top sites [Fig. 4(o)]. As the amount of S in the structure increases ($Mo-S_2$ and $Mo-S_3$), the surface mobility progressively enhances without embedding [Fig. S13]. This finding indicates that providing an excess of S is critical to preventing the sinking of bare metal atoms and subsequent exchange. To deeply understand the dynamic behaviour of Mo atoms and their clusters with S atoms on various TMD substrates, we conducted further MLP-MD simulations to investigate the migration mechanism and stability of Mo clusters on $WS_2$ surfaces. We found that their behaviour was similar to that observed on $MoS_2$ substrate (Fig. S14). Figures S15 and S16 illustrate the behaviour of single Mo atoms and Mo-S clusters on $MoS_2$ and $WS_2$ surfaces at a lower temperature (900 K). The results indicate that, even at reduced temperatures, single Mo atoms still tend to rapidly embed into the substrates. However, the surface mobility of Mo-S and $Mo-S_2$ clusters is significantly reduced compared to their behaviour at 1100 K. In contrast, $Mo-S_3$ clusters maintain relatively high surface stability and mobility, further validating the critical role of sulfur-containing clusters in suppressing Mo atom sinking [Fig. S15, Fig. S16]. Figure S17 further confirms the reliability and accuracy of these MLP-MD results through AIMD simulations at 1100 K for 10 ps. Although the simulation duration was shorter due to computational limitations of AIMD, the clusters displayed similar characteristics as observed in the MLP-MD simulations. Furthermore, these sulfur-rich structures exhibit faster surface diffusion, which is beneficial for the nucleation and aggregation of TMD layers, thereby accelerating defect healing. Experimentally, the sulfur source and molybdenum source ratios used during the MOCVD growth of $MoS_2$ and $WS_2$ are significantly higher than 2:1, such as 70:1[38], 660:1[39], 6400:1[40], and 11111:1[41]. Similarly, regarding the successful synthesis in the above-mentioned study[23], we propose that sulfur-rich conditions likely played a decisive role. Since the Mo film is deposited after the synthesis of the first $WS_2$ layer, residual sulfur remaining in the growth chamber or on the surface from the preceding step is inevitable. This residual sulfur can react with deposited Mo atoms to form Mo-S clusters during the deposition or the initial heating phase, effectively suppressing the embedding of bare metal atoms and the formation of the SMMS intermediate, as predicted by our simulations. This mechanism explains how alloying is prevented in the experimental two-step process.

By co-depositing Mo and S atoms to form MoS clusters, the homogeneous epitaxy and heteroepitaxy of the second layer of $MoS_2$ on $MoS_2$ and $WS_2$ were simulated at 1100 K, respectively. These simulations are analogous to the MOCVD growth of $MoS_2$ [see Supplementary Movies 6-7]. As shown in Fig. S18(c), the simulation starts by placing a triangular 1H-$MoS_2$ nucleus on the surface of the monolayer 1H-$MoS_2$, representing the nucleation at the onset of growth. The Mo and S atoms required for the growth of one layer of $MoS_2$ were deposited onto the surface within 516 ps, after which the deposition was stopped. Figures S18(d-f) display the system from 106 ps to 400 ps; the $MoS_2$ layer exhibits characteristics of the 1T phase. After 2 ns of simulation, the growth of the second layer of $MoS_2$ on the monolayer $MoS_2$ is essentially complete, as shown in Fig. S18(g). The 1T phase present in the early stages diminishes, indicating that $MoS_2$ transitions from the 1T phase to the more stable 1H phase as growth proceeds. The phase transition is accompanied by the healing of structural defects. The simulation of growing the second layer of $MoS_2$ on a monolayer $WS_2$ proceeds similarly, as shown in Fig. S18(h-l), producing a growth pattern characteristic of the 1T phase [Fig. S18(j-k)]. As the simulation progresses further, the 1H phase of $MoS_2$ eventually forms on the $WS_2$ substrate, as shown in Fig. S18(l).

## Electronic Properties and Device Potential of Intermediate Structures

As previously discussed, the SMMS intermediate structure impedes the growth of non-alloyed $MoS_2/WS_2$ vdWHs. Fig. 5(a) and Fig. 5(b) show the electronic band structures. Both SMoMoS and SMMS exhibit metallic properties.

We further investigated the contact characteristics of these metallic intermediate structures with the semiconductor $MoS_2$. In both cases, the contact between the metallic intermediate structures and the semiconductor $MoS_2$ forms a p-type Schottky contact, as shown in Fig. 5(c) and Fig. 5(d). The Schottky-Mott limit predicts p-type SBHs of 0.55 eV for SMoMoS and 0.69 eV for alloyed SMMS. For conventional metal electrodes (e.g., Ti, Cr, Au, Pd) interfaced with $MoS_2$, the predicted SBH ranges from 0.56 eV to 1.86 eV [42]. However, Fermi-level pinning induces strong n-type behaviour at $MoS_2$ interfaces. Experimentally observed p-type SBHs range from 1.56 to 1.75 eV. We constructed SMMS-$MoS_2$ and SMoMoS-$MoS_2$ interfaces. Compared to Schottky-Mott predictions, we observed significantly reduced Fermi-level pinning (Fig. S19). The p-type SBH increased by only approximately 0.1 eV in the same theoretical framework. Moreover, continuous SBH tuning can be achieved by adjusting metal deposition parameters, enabling tailored contact properties for electronic or optoelectronic applications. These results suggest that SMMS and SMoMoS are promising candidates for $MoS_2$ FET electrodes.

A very recent experimental study[43] realized an atomic layer bonding (ALB) contact by establishing a metallic coherent bonding interface between the transition-metal layer of TMDs and metal electrodes. This ALB structure exhibits direct metal-metal bonding, consistent with the observations in the predicted SMMS intermediate. Consistent with our findings that SMMS exhibits metallic character and improved contact properties, the experimental ALB contact demonstrated ultralow contact resistance and high thermomechanical stability. This strongly corroborates our theoretical prediction that establishing coherent metal-metal interactions (as in SMMS) is an effective strategy to overcome the limitations of van der Waals contacts and achieve high-performance electronic devices.

## Discussion

In summary, we employed MLP-MD simulations to investigate the growth mechanisms of TMD vdWHs, with a particular focus on the heterostructures of $MoS_2$ and $WS_2$. Our study revealed the stability and transformation mechanisms of the intermediate structures SMoMoS and SMMS formed during growth. Through detailed energy and kinetic stability analysis, we observed that these intermediate structures exhibit significant stability under specific conditions and transition to bilayer TMD structures upon further deposition of S atoms. Additionally, we studied the contact characteristics of these metallic intermediate structures with the semiconductor $MoS_2$ and discovered that they exhibit low *p*-type SBHs, indicating their potential applications in electronic devices. These findings provide critical theoretical insights into the growth mechanisms of TMD vdWHs and optimization of growth conditions while offering further perspectives for designing 2D materials and devices.

## Methods

### DFT

DFT calculations were performed using the Vienna Ab initio Simulation Package (VASP) [44,45] version 6.3.0. A plane wave basis set was employed, utilizing the projector augmented wave (PAW) method and standard pseudopotentials. The Perdew-Burke-Ernzerhof (PBE) [46] exchange-correlation functional within the Generalized Gradient Approximation (GGA) [47] was employed to compute the electronic structure and energy. The DFT-D3(Becke-Johnson) [48,49] van der Waals correction was chosen to account for dispersion interactions. A plane-wave cutoff energy of 500 eV and no symmetry constraints were applied. To ensure accuracy, the electronic self-consistent loop was converged to a tolerance of $10^{-5}$ eV. Gaussian smearing with a width of 0.05 eV was employed to facilitate the convergence of calculations. Spin-polarized calculations were performed. For all periodic structures, a centre-symmetric k-point mesh with a density of 0.3 $Å^{-1}$ was employed. The electronic band structure calculations were enhanced through the implementation of the HSE06 hybrid functional for both band structures and electrostatic potentials. A Fermi-Dirac smearing parameter of 0.2 eV was employed in the calculations. MD simulations were conducted using the on-the-fly machine learning potential from the VASP package, extracting frames from DFT calculations as data to accelerate the construction of the initial training set.

### MLP

The MLP was trained using the NequIP [24] framework, which implements E(3)-equivariant graph neural networks. This network provides enhanced stability for MD simulations relative to other networks, enabling extended stable simulations. The model utilized 12 radial basis functions and a maximum angular momentum of 2 for the interatomic edges. The hidden layers included irreducible representations as follows: 128×0e, 64×0o, 128×1o, 64×1e, 32×2o, and 32×2e. Element information was embedded as 128-dimensional vectors. Five graph convolution layers were implemented, including self-connections and residual connections. The cutoff distance for constructing the graph was set at 6 Å. The loss function included both energy per atom and atomic force terms, each contributing equally to the total loss. Each loss term used mean squared error. During the initial training, we used a learning rate of 0.005 and a batch size of 5. For the iterative learning process, we employed a lower learning rate of $10^{-4}$ to fine-tune the model. To enhance the model's robustness, we added a constant repulsive term to the output layer:

$$E_{\text{repulsive}} = \left(\frac{r_0}{r}\right)^{12} \frac{r_0}{24} \text{cutoff}(r)$$

This approach enhanced the model's stability, with $r_0 = 1.8$ Å in this study. The iterative learning process allowed us to continuously refine the model by incorporating new structures and configurations, ensuring comprehensive coverage of the complex structural landscape involved in TMD heterostructure growth.

## MD

MD simulations were conducted using the Large-scale Atomic/Molecular Massive Parallel Simulator (LAMMPS) [50], utilizing our developed MLP. We employed the bin algorithm for neighbour list construction with a cutoff radius of 6.0 Å and a skin distance of 2.0 Å. The neighbour list was rebuilt only when at least one atom had moved beyond half of the skin distance threshold. Simulations were conducted within the canonical ensemble (NVT) employing a Nosé-Hoover chain thermostat [51,52], with a temperature damping time set to 0.1 ps. An integration time step of 1.0 fs was employed to ensure performance while maintaining simulation stability. Initial velocities were sampled from a Gaussian distribution. The degrees of freedom contributing to the system temperature were dynamically updated with the deposition of new atoms. The simulations employed an ideal Morse substrate ( $V(r) = D_0\left(1 - e^{-\alpha(r-r_0)}\right)$ ) with parameters $D_0 = 0.2$ eV, $\alpha = 1.5$, and $r_0 = 3.5$ Å. This wall mimics the weak van der Waals interaction (< 0.1 eV/atom) between Mo/W atoms and oxide substrates such as sapphire, thus providing only mechanical support while avoiding spurious interfacial chemistry. For Mo atom deposition, we used a $4\sqrt{3} \times 7$ $MoS_2$ or $WS_2$ supercell at 900 K. This temperature was chosen to represent the thermal state during the pre-heating/baking stage and the ramp-up process prior to full sulfurization. The SMoMoS and SMMS sulfurized process simulations were performed using a $4\sqrt{3} \times 7$ supercell at 1100 K, aligning with the experimental sulfurization temperatures (approx. 1073 K). The diffusion behaviour of Mo-S clusters on the $MoS_2$ surface was also investigated using a $4\sqrt{3} \times 7$ $MoS_2$ supercell at 1100 K. For the simultaneous Mo and S deposition to grow the second $MoS_2$ layer, we employed a larger $6\sqrt{3} \times 8$ $MoS_2$ or $WS_2$ supercell at 1100 K. The atomic deposition was achieved through the fix deposit command in LAMMPS, which introduced a Mo atom with a downward velocity of 5 Å/ps at random positions within the top region of the simulation box every 6000 steps. The initial kinetic energy of deposited atoms was set to 0.12 eV. To ensure the reproducibility of the non-equilibrium processes, all key qualitative observations (including the sinking of Mo atoms and the sulfurization-induced extraction) were confirmed in three independent MD simulations initialized with different random velocity seeds.

## Data availability

The training data, trained models, and computation setting files have been uploaded to the Zenodo repository (https://doi.org/10.5281/zenodo.18397127). Source data are provided with this paper.

## Code availability

The training program has been uploaded to a public GitHub repository (https://github.com/1713175349/ocp).

## Acknowledgements

The authors acknowledge the financial support provided by the National Key R&D Program of China (2024YFA1409600 to J.G.), R&D project of Joint Funds of Liaoning Province [2023JH2/101800038 to J.G.], the Dalian Science and Technology Innovation Fund (2025JJ12GX012 to J.G.), the National Natural Science Foundation of China (12374253 to J.G., 12374174 to H.L., 12504315 to Y.C.), and the National Foreign Expert Project (D20240213 to J.G., D20240220 to H.L.). Gao thanks Z G Yu (A*STAR) for great helpful discussions and suggestions. The authors also acknowledge Computers supporting from Shanghai Supercomputer Center.

## Contributions

J.G.: Conceptualization, Supervision, Funding acquisition, Writing – review & editing. L.Z.: Methodology, Investigation, Formal analysis, Writing – original draft. F.D.: Supervision, Writing – review & editing. H.L.: Funding acquisition, Writing – review & editing. Y.C., X.S., J.Z.: Writing – review & editing.

## Competing Interests

The authors declare no competing interests.

## Figure Legend

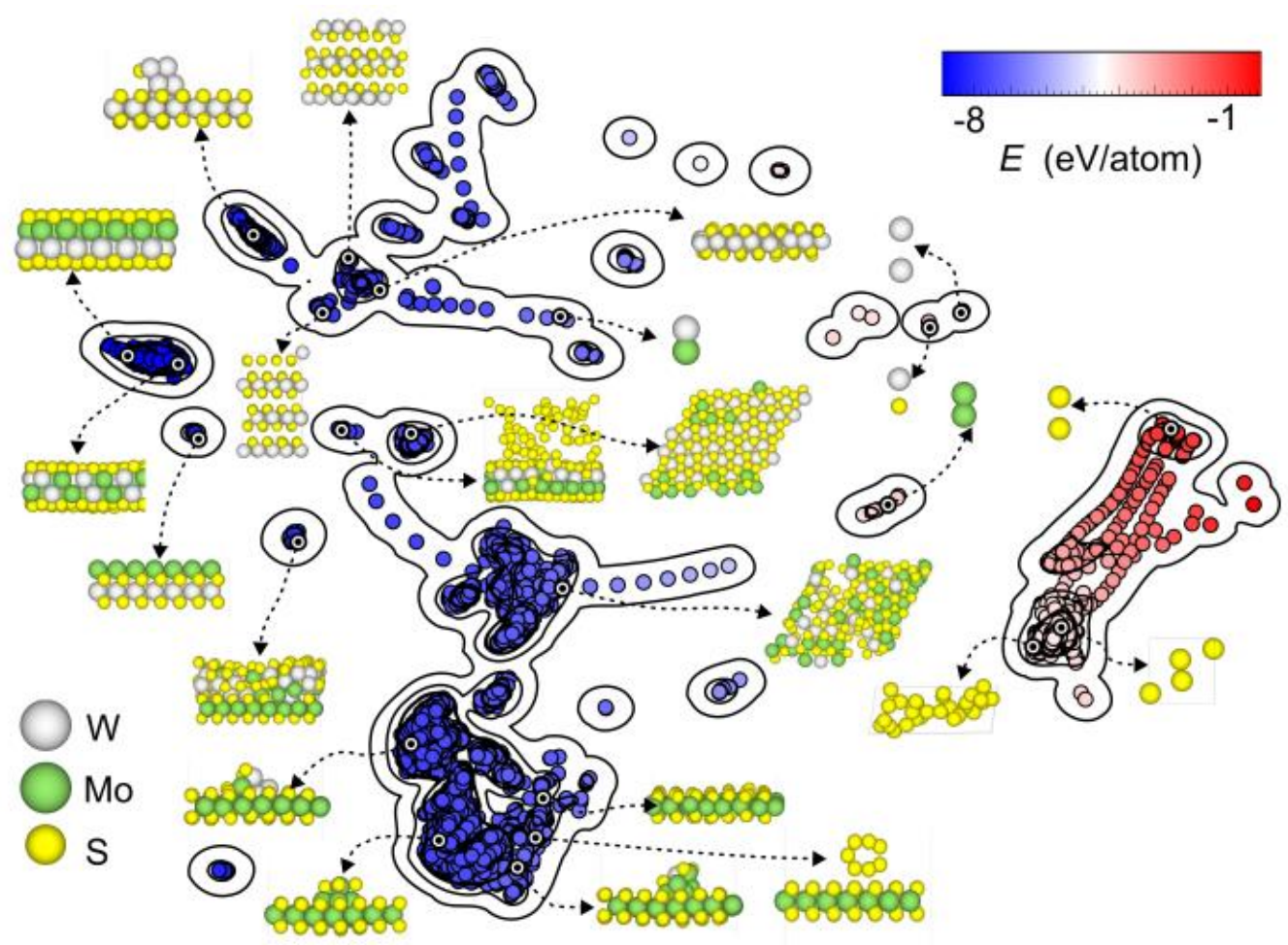

**Figure 1. Structural diversity of the dataset.** The key structures within the dataset are visualized using principal component analysis (PCA) to illustrate the diversity of TMDs, $MoS_2/WS_2$ vdWHs, and various complex structures that emerge during growth processes. The density of dataset in specific areas is represented by contour lines. Source data are provided as a Source Data file.

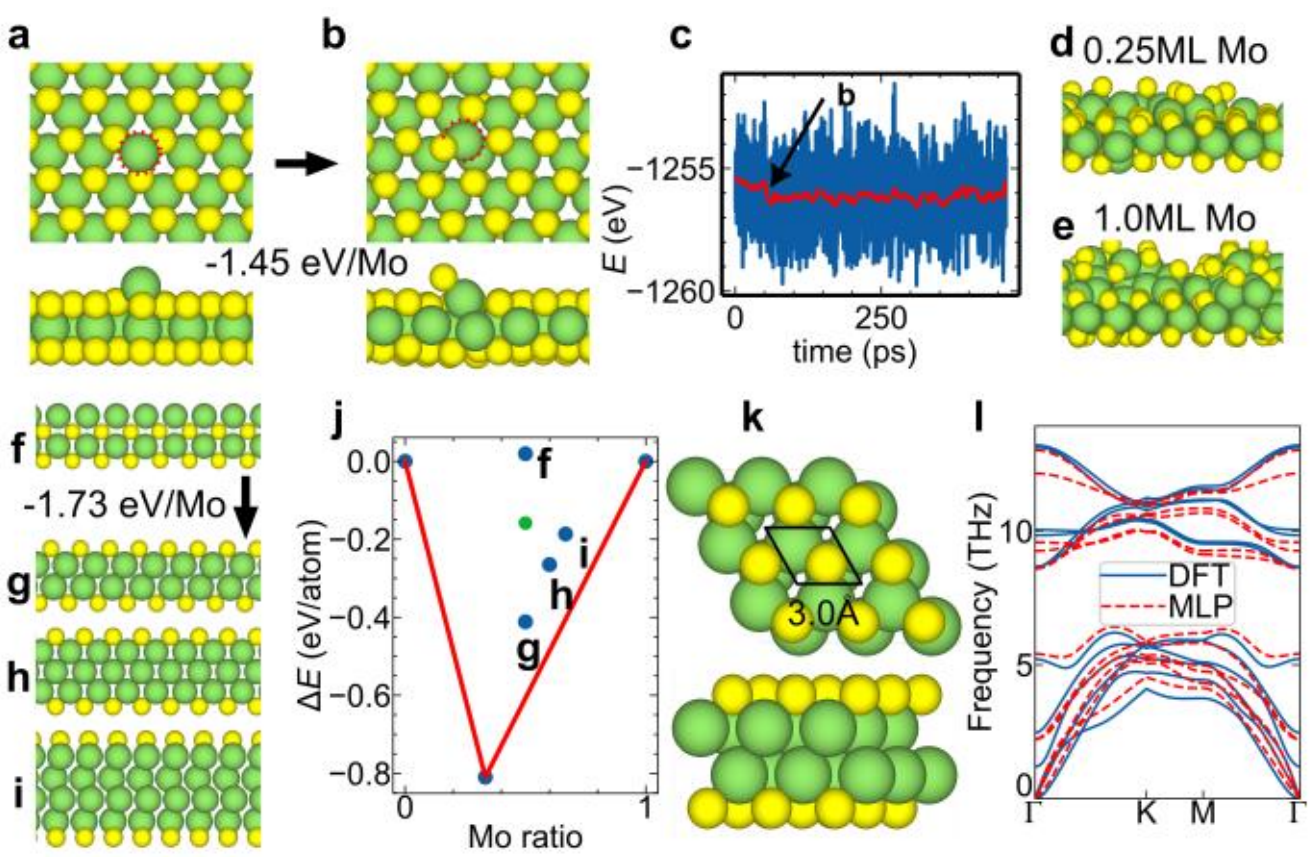

**Figure 2. Deposition dynamics and stability of Mo atoms on $MoS_2$.** (a) The structure of Mo atoms deposited and (b) embedded into the $MoS_2$ layer, (c) energy evolution during the MLP-MD simulation at 1100 K (blue line represents raw energy, while red line shows the low-pass-filtered result). (d-e) Snapshots of 0.25 monolayer (ML) (d) and 1.0 ML (e) Mo atom deposition simulated at 900 K on the existing $MoS_2$ layer. (f-i) Four possible structures: (f) MoSMoS, (g) SMoMoS, (h) $SMo_3S$, and (i) $SMo_4S$ with more Mo atoms embedded. (j) Formation energy convex hull of the considered MoS structures, with green dots corresponding to energies of structures found in the 2D Materials Database (MatHub-2d). (k) The unit cell structure of the SMoMoS and (l) its phonon dispersion relation. In panels (a, b, d, e, f-i, k), the green and yellow spheres represent Mo and S atoms, respectively. Source data for (c, j, l) are provided as a Source Data file.

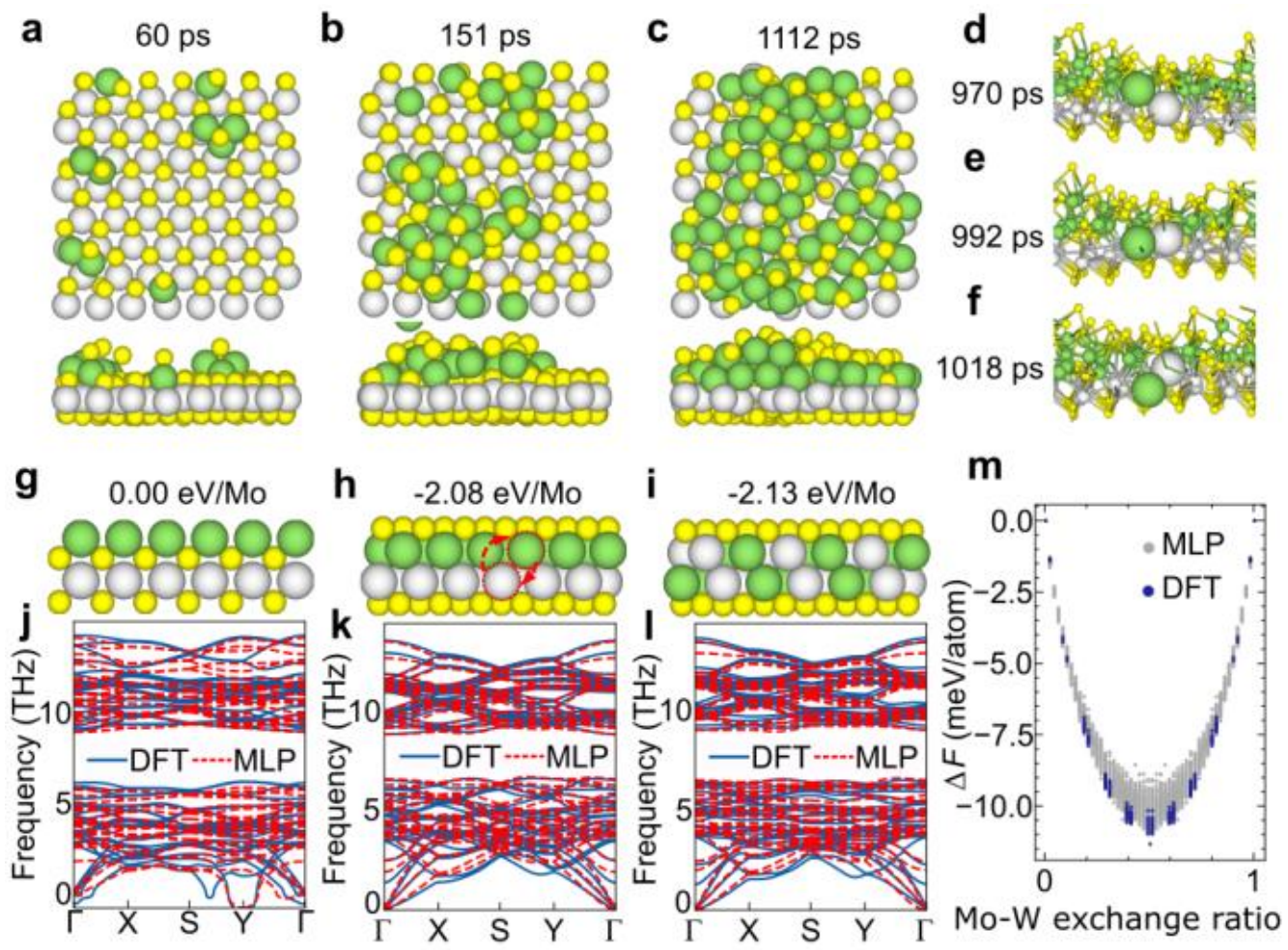

**Figure 3. Heterogeneous deposition dynamics and alloying of Mo atoms on $WS_2$.** (a-c) Snapshots of the growth structure of $MoS_2/WS_2$ vdWHs during the two-step vapor-deposition process by MLP-MD simulation (Mo atoms are deposited on $WS_2$) (900 K). (d-f) Observed exchange phenomena of Mo and W atoms during the MLP-MD simulation. (g) Schematic of the Mo layer on the $WS_2$ surface and (j) related phonon dispersion. (h) Schematic of the unalloyed SMoWS intermediate structure and (k) related phonon dispersion. (i) Schematic of the alloyed SMMS structure and (l) related phonon dispersion. (m) Relative free energy ($\Delta F$) of different Mo-W exchange ratios of the SMMS structure at 300 K. In panels (a-i), the green, grey-white, and yellow spheres represent Mo, W, and S atoms, respectively. Source data for (j-m) are provided as a Source Data file.

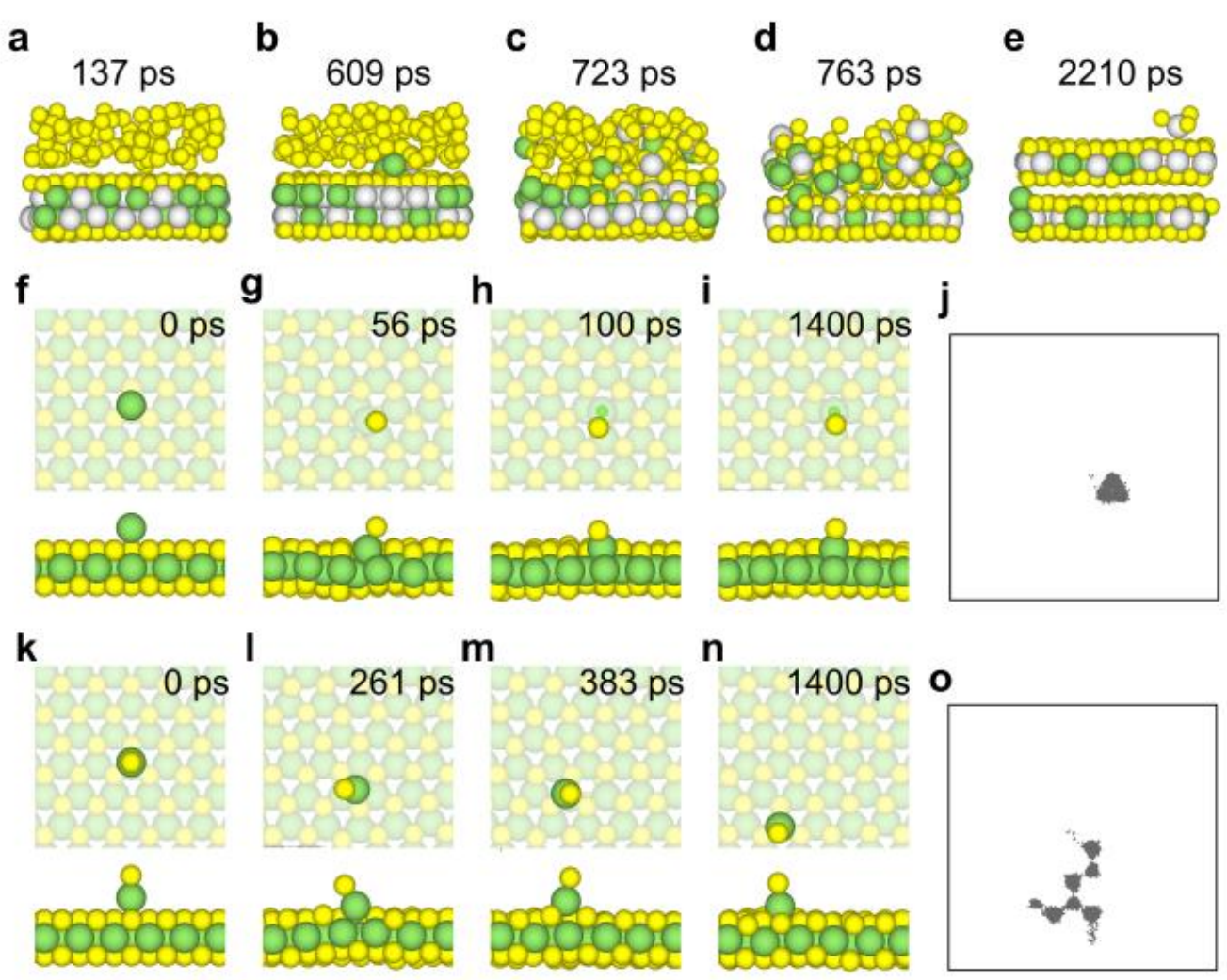

**Figure 4. Sulfurization dynamics and surface behaviour of Mo-S clusters.** (a-e) MLP-MD simulations of the growth of alloyed $Mo_xW_{1-x}S_2/Mo_{1-x}W_xS_2$ vdWHs by depositing S atoms on the alloyed SMMS intermediate phase (1100 K). (f-o) MLP-MD simulations of various Mo clusters on $MoS_2$ (1100 K): (f-i) deposition of single Mo atoms; (j) trajectories of Mo atoms in the $xy$ plane; (k-n) deposition of Mo-S clusters and (o) trajectories of Mo atoms in the $xy$ plane. In panels (a-i, k-n), the green, grey-white, and yellow spheres represent Mo, W, and S atoms, respectively. Source data for (j, o) are provided as a Source Data file.

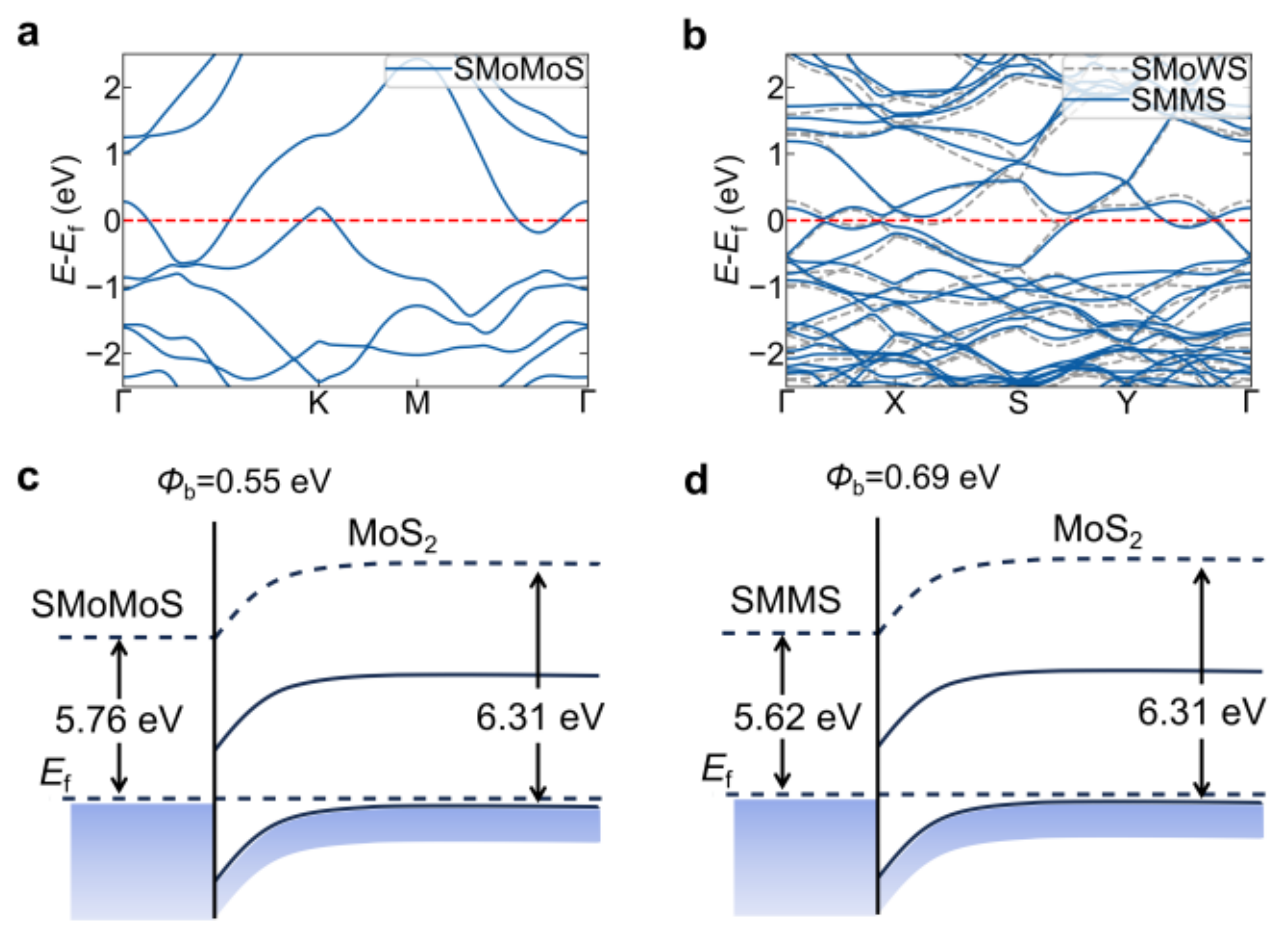

**Figure 5. Electronic structures and contact properties of SMoMoS and SMMS electrodes with $MoS_2$.** (a) Electronic band structure of the SMoMoS structure ($E_f$ is the Fermi level). (b) Electronic band structures of the SMoWS structure and alloyed SMMS structure ($E_f$ is the Fermi level). (c) Schematic of the p-type Schottky barrier (SBH) at the $MoS_2$-SMoMoS interface, with $\Phi_b$ = 0.55 eV. (d) Schematic of the p-type SBH at the $MoS_2$-SMMS interface, with $\Phi_b$ = 0.69 eV. Source data for (a, b) are provided as a Source Data file.

Supplementary Information for

# Intermediates of Forming Transition Metal Dichalcogenide Heterostructures Revealed by Machine Learning Simulations

Luneng Zhao, Hongsheng Liu, Yuan Chang, Xiaoran Shi, Jijun Zhao, Feng Ding and Junfeng Gao*

Corresponding authors: Junfeng Gao, gaojf@dlut.edu.cn

The supplementary materials provide further details to substantiate the findings presented in the main manuscript. We include assessments of the precision of the machine learning potential (MLP) developed for our simulations, extended discussions on the unique intermediate structures identified during transition metal dichalcogenide (TMD) growth, and additional simulation results exploring the kinetics of metal atom embedding and exchange. Lastly, we discuss potential applications of the SMMS intermediate structure in electronic devices.

## 1. Validation of MLP Accuracy

In this study, we employed an iterative training approach to develop the MLP. The process began with the generation of an initial dataset using ab initio molecular dynamics (AIMD) simulations. Subsequently, we assessed the model's reliability by benchmarking it against DFT calculations and applying it to simulate the two-step growth of $MoS_2/WS_2$. If the MLP exhibited instabilities during molecular dynamics or failed to reproduce accurate phonon spectra, we refined the model by incorporating additional relevant DFT configurations into the training set. This cycle of data expansion and model retraining was repeated iteratively until the MLP achieved satisfactory accuracy and stability. The final optimized model was then deployed for efficient and accurate large-scale atomistic simulations.

To generate the initial training set, we employed on-the-fly molecular dynamics (MD) simulations, a method that effectively explores unknown configuration spaces while minimizing sampling costs. These simulations were conducted using the VASP on-the-fly MD framework with parameters optimized for efficiency and accuracy. Specifically, we utilized a timestep of 3.5 fs to capture atomic motions precisely, with a maximum of 20,000 steps per trajectory. The simulations were performed in the canonical (NVT) ensemble using Nosé-Hoover thermostats, covering a temperature range from 500 K to 1300 K to encompass typical growth conditions. The unit-cell shapes were kept fixed, and the number of local reference configurations in the on-the-fly MLP was constrained by setting ML_MB=2000. The initial configurations included monolayer $MoS_2$ and $WS_2$, as well as their combinations featuring homogeneous/heterogeneous junctions and adsorbed metal clusters. Subsequently, the training set was expanded through iterative training rounds employing distinct sampling strategies:

• **Training round 1:** On-the-fly MD sampling of MSMS structure dynamics.

• **Training round 2:** Randomized metal-sulfur exchange sampling — Starting from the relaxed MSMS structure, we performed random substitutions between metal (Mo/W) and sulfur atoms at specific interfacial sites to generate intermediate SMMS-like configurations. These exchanges emulate early-stage metal–chalcogen mixing observed during the deposition process.

• **Training round 3:** Randomized metal-metal swapping — To construct a diverse set of alloyed SMMS structures, we performed layer-constrained Mo-W swaps across top and bottom layers. This preserves the overall composition but introduces configurational disorder relevant to alloying behaviour.

• **Training round 4:** Surface metal-sulfur exchange sampling — Starting from SMMS structures, we randomly exchanged surface S atoms with underlying metal atoms (Mo or W) to simulate possible sulfurization pathways and local rearrangements during the growth process.

• **Training round 5:** Perturbed configurations near SMMS equilibrium states.

• **Training round 6:** On-the-fly MD sampling of sulfur molecular dynamics (matching initial dataset generation method).

• **Training round 7:** S-S/Mo-S/Mo-W dimer configurations at varied distances.

• **Training rounds 8-17:** MLP-MD simulations of SMMS growth and Mo deposition (key structures extracted every 10,000 steps).

During each training round, we fine-tune the MLP model through 1:1 random sampling of historical data (cumulative from prior training rounds) and newly generated data. The MLP fine-tuning employed a learning rate of $1\times10^{-4}$ to balance convergence speed and accuracy. The final dataset contained approximately 26,000 configurations from initial training and all training rounds, ensuring structural diversity and representativeness critical for robust MLP development. The dataset has been uploaded to the **Zenodo repository (https://doi.org/10.5281/zenodo.18397127)**.

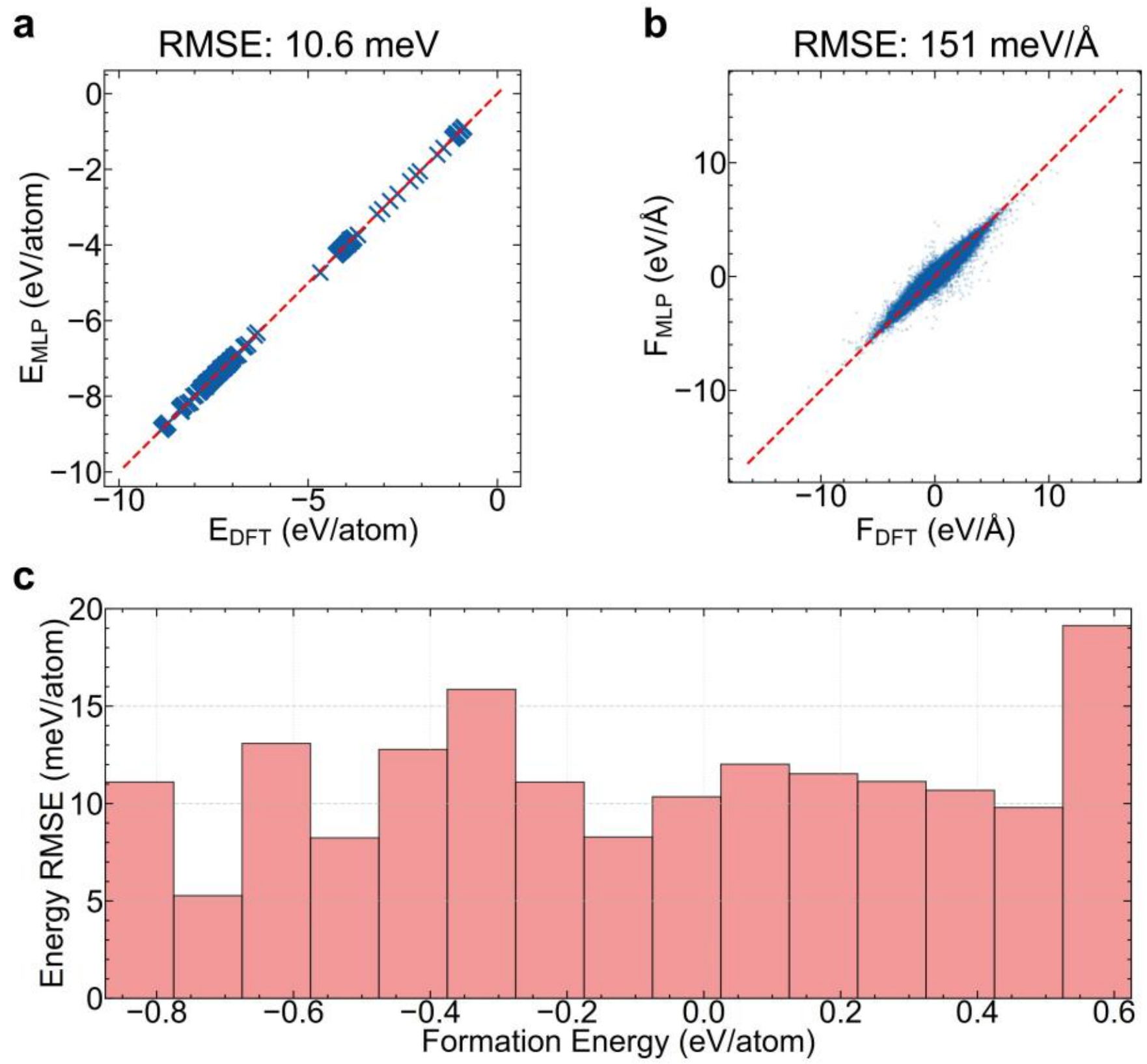

**FIG. S1:** Comparison of MLP predictions with DFT calculations. (a) Correlation between MLP-predicted energies and DFT-calculated energies per atom. The root mean square error (RMSE) is 10.6 meV/atom. (b) Correlation between MLP-predicted forces and DFT-calculated forces. The RMSE is 151 meV/Å. The red dashed lines represent perfect correlation. (c) Distribution of energy RMSE for test set structures across varying formation energy intervals. The histograms display the prediction errors for energy in meV/atom. The formation energy per atom determines the binning interval.

To assess the MLP's accuracy and reliability for simulating TMD heterostructure growth, we reserved 10% of the overall dataset as a test set. FIG. S1 presents a comparison between MLP predictions and DFT calculations. As shown in FIG. S1(a), it demonstrates the MLP's precision in predicting atomic energies, with a RMSE of 10.6 meV/atom [1]. FIG. S1(b) demonstrates the MLP's precision in predicting atomic forces, with a RMSE of 151 meV/Å. FIG. S1(c) provides a more detailed error analysis by categorizing the test structures based on their formation energy per atom. The histogram displays the energy RMSE distribution across different formation energy intervals. This analysis reveals that the model maintains exceptional accuracy for structures in the lower energy regimes (typically corresponding to equilibrium or near-equilibrium configurations). Crucially, even for higher-energy states, which represent distorted configurations or transition states frequently encountered during high-temperature growth simulations, the energy prediction errors remain low and bounded. This comprehensive validation confirms the MLP's robustness and consistent accuracy across the entire relevant energy landscape.

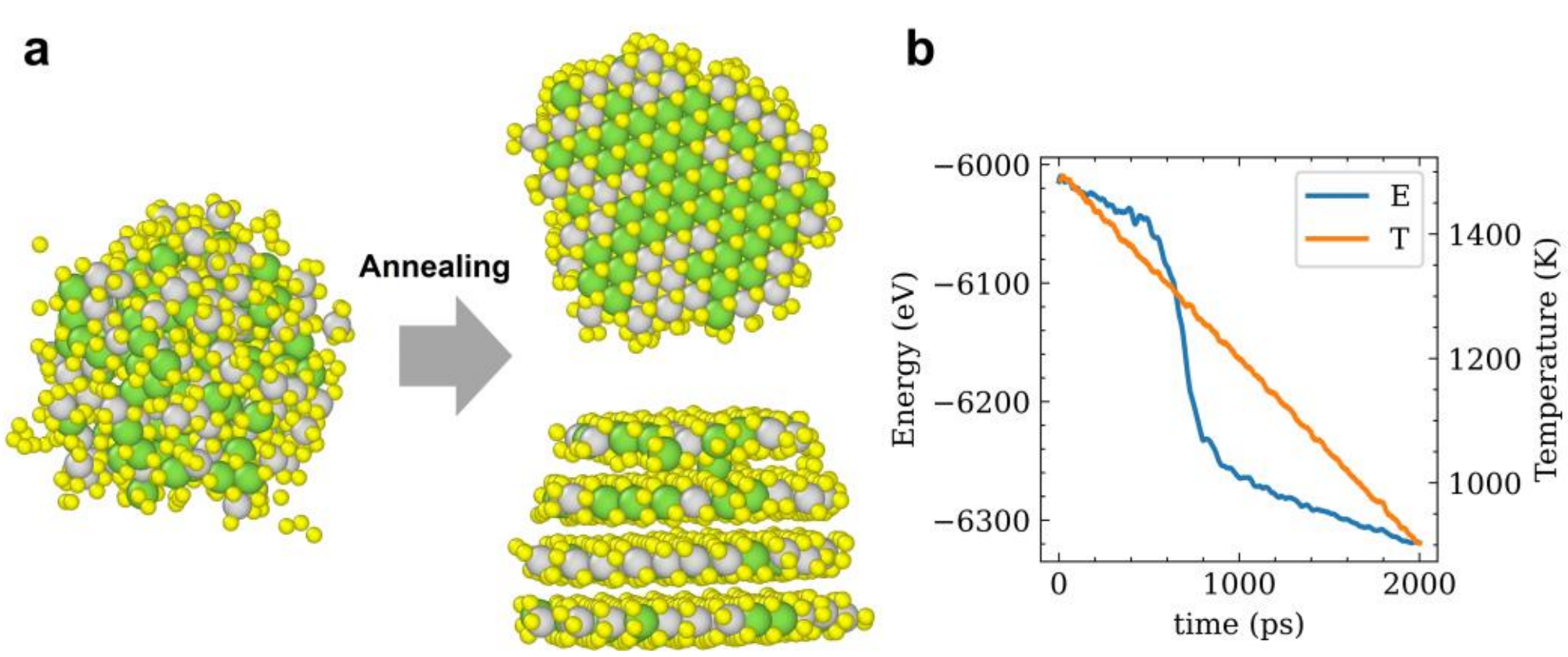

**FIG. S2:** Formation of ordered TMD layers through annealing simulation. (a) Snapshots depicting the evolution from an initial random mixture of Mo, W, and S atoms (left, ratio 1:1:4) to ordered TMD layers after annealing. (b) Time evolution of system energy (E, blue line) and temperature (T, orange line) during the 2 ns annealing process, illustrating the formation of stable TMD structures. Note: The data has been smoothed using a low-pass filter for clarity.

To validate the MLP's capability in describing growth processes, we conducted simulated annealing experiments. FIG. S2 illustrates the results of these simulations. As shown in FIG. S2(a), we started with a random mixture of Mo, W, and S atoms (left image) and observed their evolution into ordered TMD layers after annealing (right images, showing top and side views). This transformation demonstrates the MLP's ability to accurately capture the self-assembly process of TMD structures.

FIG. S2(b) presents the evolution of the system energy (E, blue line) and temperature (T, orange line) during the 2 ns annealing process. The decreasing energy trend, coupled with temperature fluctuations, indicates the formation of stable TMD structures. It's worth noting that a low-pass filter was applied to smooth the results, enhancing the clarity of the overall trends. These annealing simulations validate our MLP's effectiveness in modelling TMD growth.

**Table S1. Comparison of energies (eV/atom) for critical structures calculated by DFT and MLP**

| Structure | DFT (eV/atom) | MLP (eV/atom) | ΔE (eV/atom) |
|---|---|---|---|
| S solid | -4.323 | -4.311 | -0.013 |
| 1H-$MoS_2$ | -7.582 | -7.583 | 0.001 |
| 1T-$MoS_2$ | -7.298 | -7.300 | 0.002 |
| 1T'-$MoS_2$ | -7.390 | -7.380 | -0.010 |
| 1H-$WS_2$ | -8.214 | -8.144 | -0.070 |
| 1T-$WS_2$ | -7.893 | -7.899 | 0.005 |
| 1T'-$WS_2$ | -8.026 | -8.017 | -0.009 |
| MoSWS | -8.375 | -8.416 | 0.041 |
| MoSMoS | -7.978 | -8.007 | 0.029 |
| SMoMoS | -8.409 | -8.382 | -0.027 |
| SMoWS | -8.895 | -8.892 | -0.002 |
| SMMS | -8.897 | -8.895 | -0.002 |
| SMoMoS (from MD) | -8.329 | -8.314 | -0.015 |
| SMoWS (from MD) | -8.804 | -8.815 | 0.011 |

**Table S2: Dataset statistics (including cell volumes)**

| Structure Category | Configurations | Avg. Atom Count | Typical Cell Size (Å) |
|---|---|---|---|
| TMD surface with W-Mo-S cluster interaction | 18396 | 155 | 22 × 22 × 30 |
| Single-layer Mo on $WS_2$ surface | 668 | 196 | 22 × 22 × 21 |
| Disordered alloyed TMD structures | 1231 | 144 | 22 × 22 × 23 |
| 1T-phase TMD | 1804 | 144 | 22 × 22 × 23 |
| Further sulfurized structures of SMMS | 360 | 280 | 21 × 21 × 21 |
| TMD nanotubes | 403 | 62 | 32 × 32 × 5 |
| Disordered SMMS structures | 900 | 196 | 22 × 22 × 21 |
| Ordered SMMS structures | 748 | 144 | 18 × 16 × 30 |
| MoSWS structures | 96 | 143 | 19 × 16 × 22 |
| Pristine mono-/bilayer $MoS_2$/$WS_2$ and heterostructures | 138 | 58 | 13 × 10 × 26 |
| Elemental sulfur | 1101 | 35 | 14 × 14 × 19 |
| Dimers | 589 | 2 | 25 × 25 × 25 |
| Total | 26434 | | |

**Table S3: Summary of simulation parameters used in each figure**

| Figure Reference | Temperature | Ensemble |
| --- | --- | --- |
| Fig. 2(a-c) | 1100 K | NVT |
| Fig. 2(d-e) | 900 K | NVT |
| Fig. 3(a-f) | 900 K | NVT |
| Fig. 4 | 1100 K | NVT |
| Fig. S4 | 900 K | NVT |
| Fig. S6 | 1100 K | NVT |
| Fig. S7(a) | 1100 K | NVE |
| Fig. S7(b-c) | 1100 K | NVT |
| Fig. S9 | 1100 K | NVT |
| Fig. S10 | 300 K | NVT |
| Fig. S11 | 1100 K | NVT |
| Fig. S13 | 1100 K | NVT |
| Fig. S14 | 1100 K | NVT |
| Fig. S15 | 900 K | NVT |
| Fig. S16 | 900 K | NVT |
| Fig. S17 | 1100 K | NVT |
| Fig. S18 | 1100 K | NVT |

## 2. Formation of SMoMoS Intermediate Structures

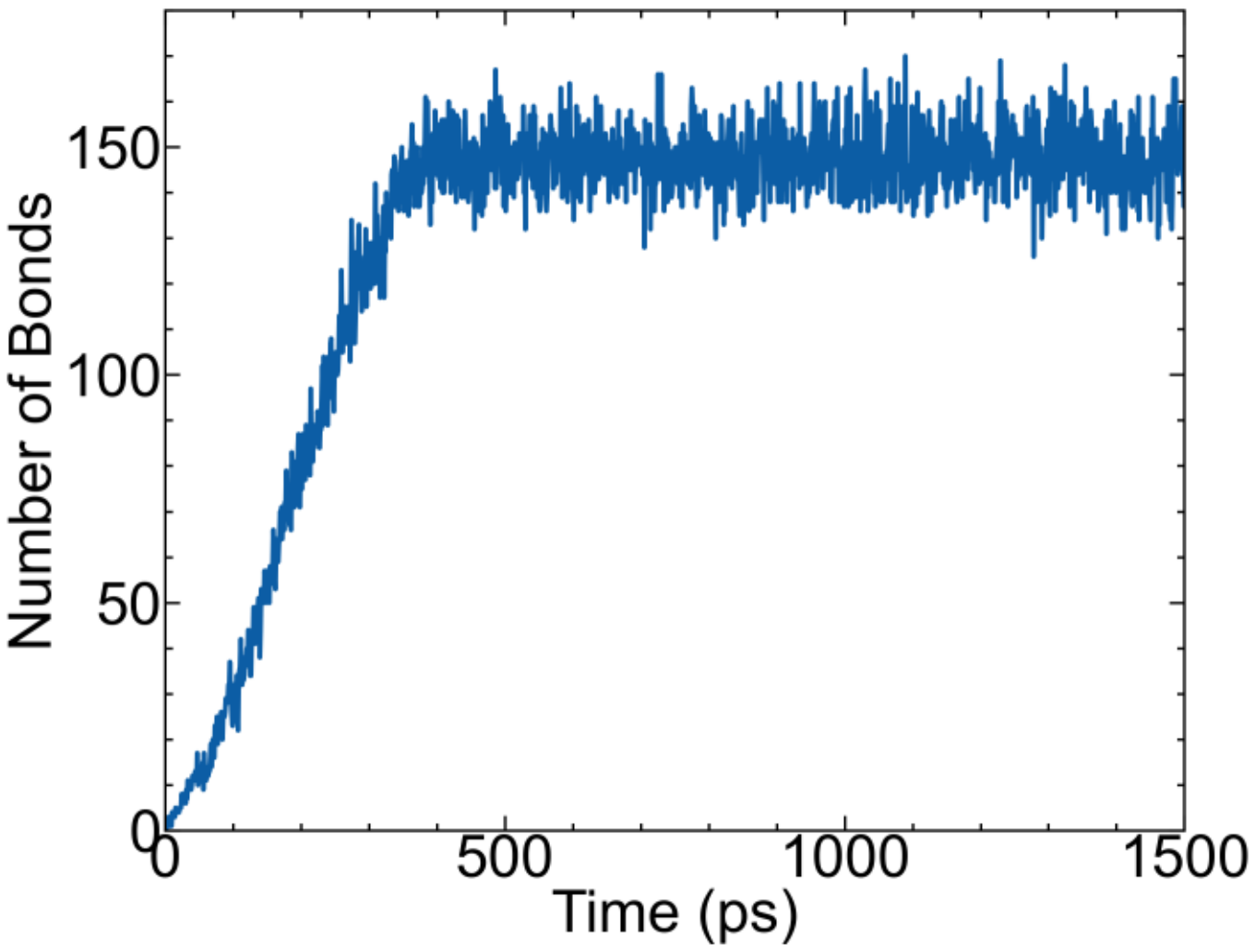

**FIG. S3:** Number of Mo-Mo bonds during Mo deposition on a $MoS_2$ substrate. Bond identification is based on a cutoff distance of 2.85 Å, which is larger than the Mo-Mo bond length in Mo bulk (~2.74 Å) but smaller than the equilibrium Mo–Mo spacing in pristine $MoS_2$ (~3.15 Å).

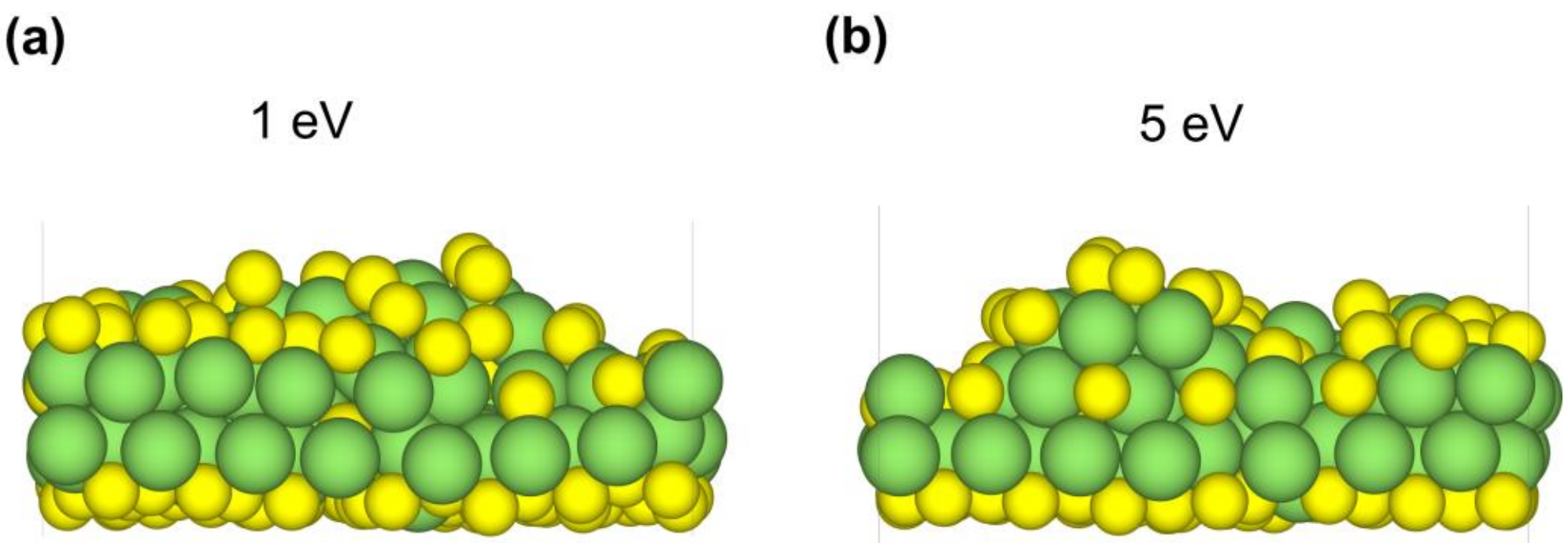

**FIG. S4:** (a) Simulated 1 eV kinetic energy deposition of a single Mo layer at 900 K on $MoS_2$. (b) Simulated 5 eV kinetic energy deposition of a single Mo layer at 900 K on $MoS_2$.

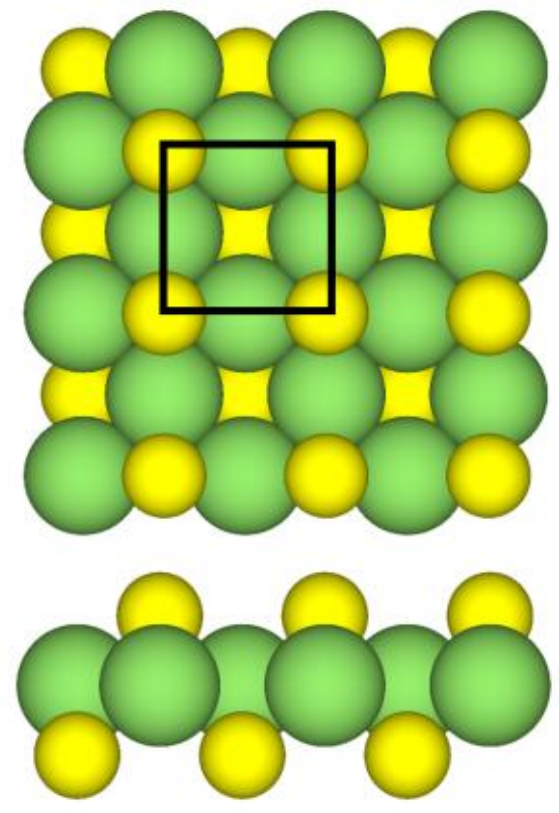

**FIG. S5:** Structure from MatHub-2d (id: MatHub2d-1817-$Mo_2S_2$).

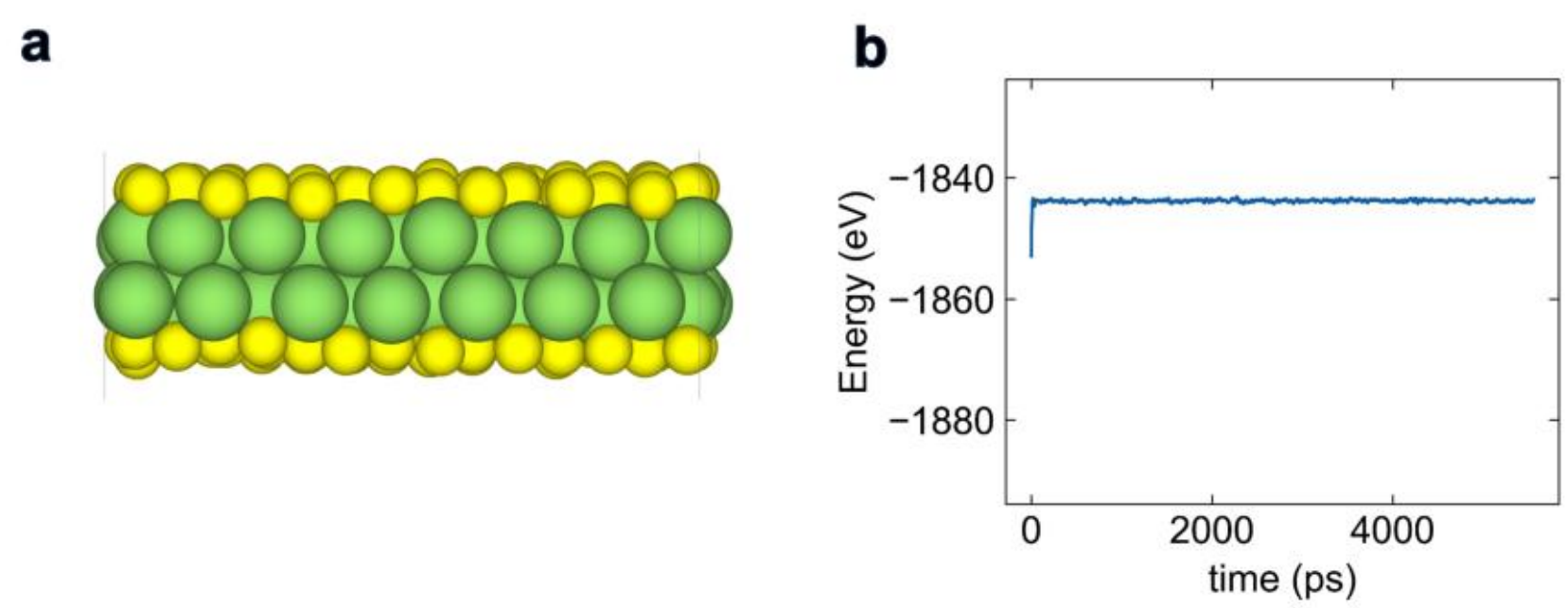

**FIG. S6:** Stability analysis of SMMS intermediate structures. (a) Atomic configurations of SMoMoS. (b) Corresponding energy evolution during 5500ps MLP-MD simulations at 1100 K for SMoMoS, demonstrating its long-term stability. Note: The data has been smoothed using a low-pass filter for clarity.

**3. Validation of the formation of SMoMoS**

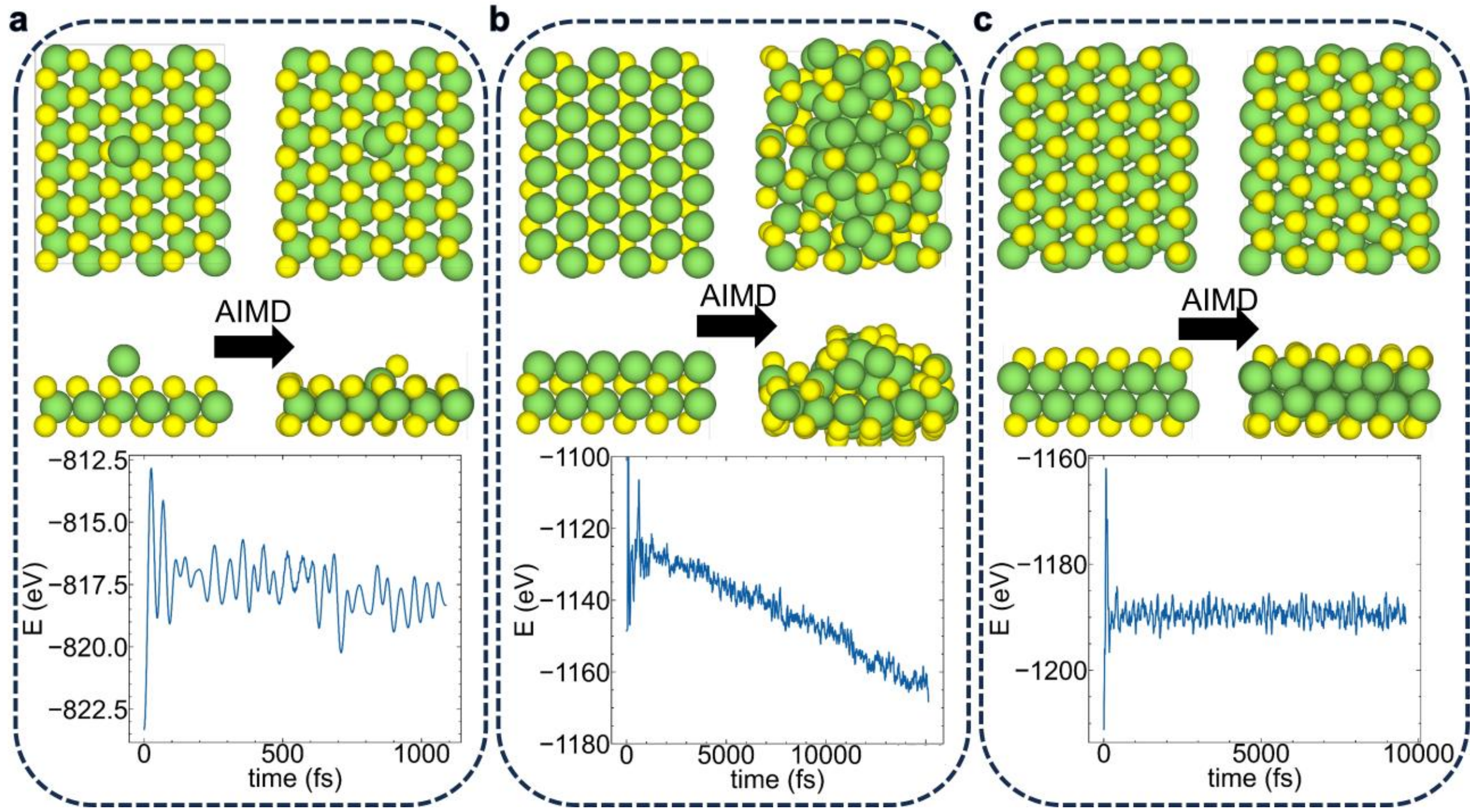

**FIG. S7:** AIMD simulations validating SMoMoS formation and stability at 1100 K. (a) Single Mo atom embedding into $MoS_2$ (~1000 fs). Top: Initial and final configurations. Bottom: Energy-time curve showing abrupt decrease at ~700 fs (NVE). (b) Mo atom layer embedding into $MoS_2$(~15 ps). Top: Initial and final states. Bottom: Energy-time curve during embedding process (NVT). (c) SMoMoS structure stability (~10 ps). Top: Initial and final states. Bottom: Energy-time curve demonstrating structural stability (NVT).

FIG. S7 presents the results of our ab initio molecular dynamics (AIMD) simulations conducted at 1100 K to validate the formation and stability of the SMoMoS structure. These simulations provide crucial insights into the behaviour of Mo atoms on $MoS_2$ surfaces and the formation mechanism of the SMoMoS intermediate structure.

FIG. S7(a) illustrates the embedding process of a single Mo atom into a $MoS_2$ layer. We initially positioned a Mo atom 2.3 Å above the $MoS_2$ surface and imparted it with a downward velocity of 5 Å/ps, corresponding to a kinetic energy of approximately 0.12 eV. This energy is comparable to the thermal energy of atoms at 1100 K. To accurately simulate thermal conditions, we sampled the initial velocities of all atoms from a Boltzmann distribution at 1100 K. The simulation was performed using the NVE (constant number of particles, volume, and energy) ensemble to capture the system's natural evolution without external temperature control. The top images show the initial and final atomic configurations, while the bottom graph depicts the energy evolution over time. Despite the relatively low initial kinetic energy, the Mo atom rapidly embeds itself into the $MoS_2$ layer within approximately 700 fs, as evidenced by the abrupt energy decrease. This embedding process releases about 1.45 eV of energy (validated by DFT calculations), demonstrating a strong thermodynamic driving force for Mo incorporation into the $MoS_2$ structure.

FIG. S7(b) extends our investigation to the behaviour of a layer of Mo atoms on $MoS_2$. The simulation runs for approximately 15 ps, allowing us to observe the long-term evolution of the system. The top images show the initial configuration with a layer of Mo atoms on the $MoS_2$ surface, and the final state where multiple Mo atoms have embedded into the $MoS_2$ layer. The bottom graph illustrates the gradual decrease in system energy as Mo atoms progressively embed into the $MoS_2$ structure. This energy reduction, significantly lower than the initial configuration, further confirms the thermodynamic favourability of Mo incorporation.

FIG. S7(c) focuses on the stability of the formed SMoMoS structure over an extended period of about 1 ps. The top images demonstrate that the SMoMoS structure remains intact throughout the simulation, with no significant structural changes observed. The bottom graph shows a stable energy curve over time, confirming the structural integrity and thermodynamic stability of the SMoMoS intermediate.

These AIMD simulations serve as a robust validation of our MLP results, confirming the accuracy and reliability of our MLP in predicting the behaviour of Mo atoms on $MoS_2$ surfaces.

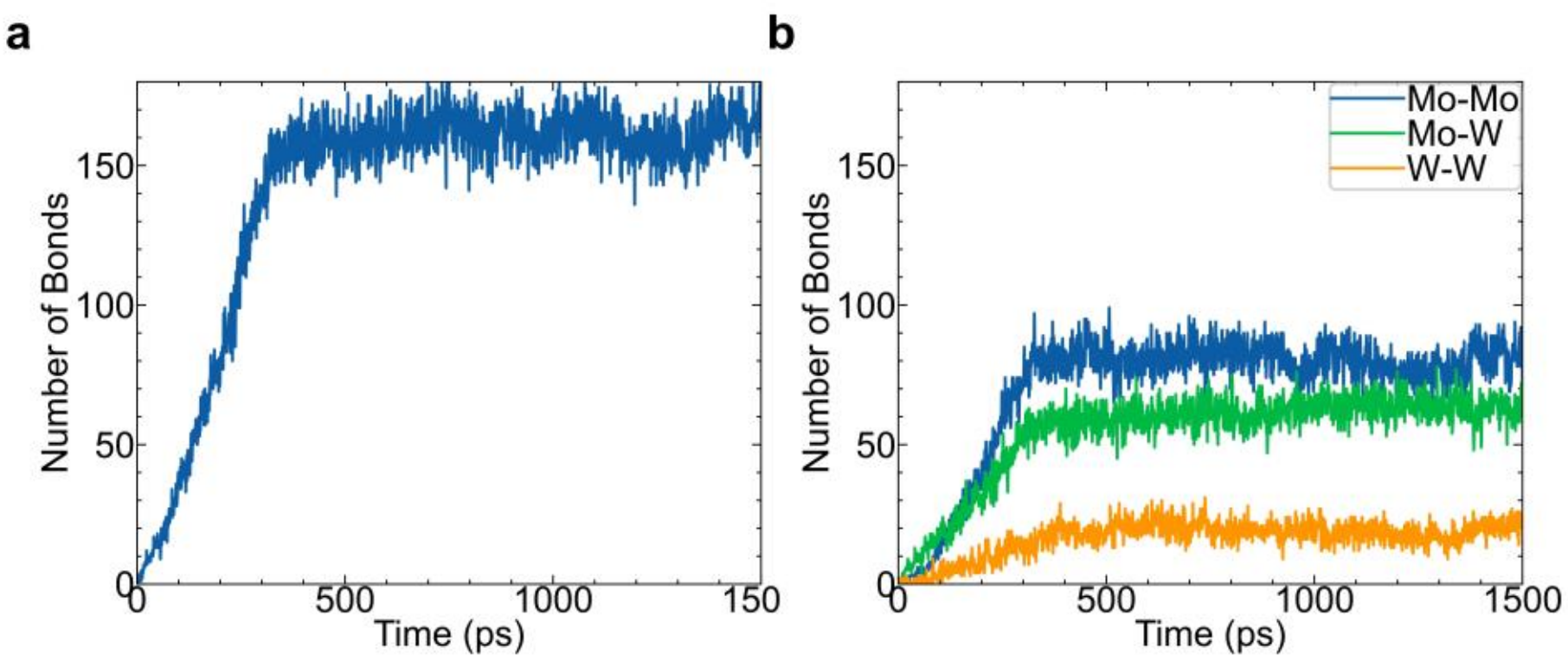

**FIG. S8:** (a) Evolution of the total number of Mo/W–Mo/W chemical bonds (including Mo-Mo, Mo-W, and W-W bonds) during Mo deposition on a $WS_2$ substrate. (b) Individual counts of Mo-Mo, Mo-W, and W-W bonds. Bond identification is based on a cutoff distance of 2.85 Å, which is larger than the Mo-Mo bond length in Mo bulk (~2.74 Å) and W bulk (~2.75 Å) but smaller than the equilibrium Mo-Mo spacing in pristine $MoS_2$ (~3.15 Å)

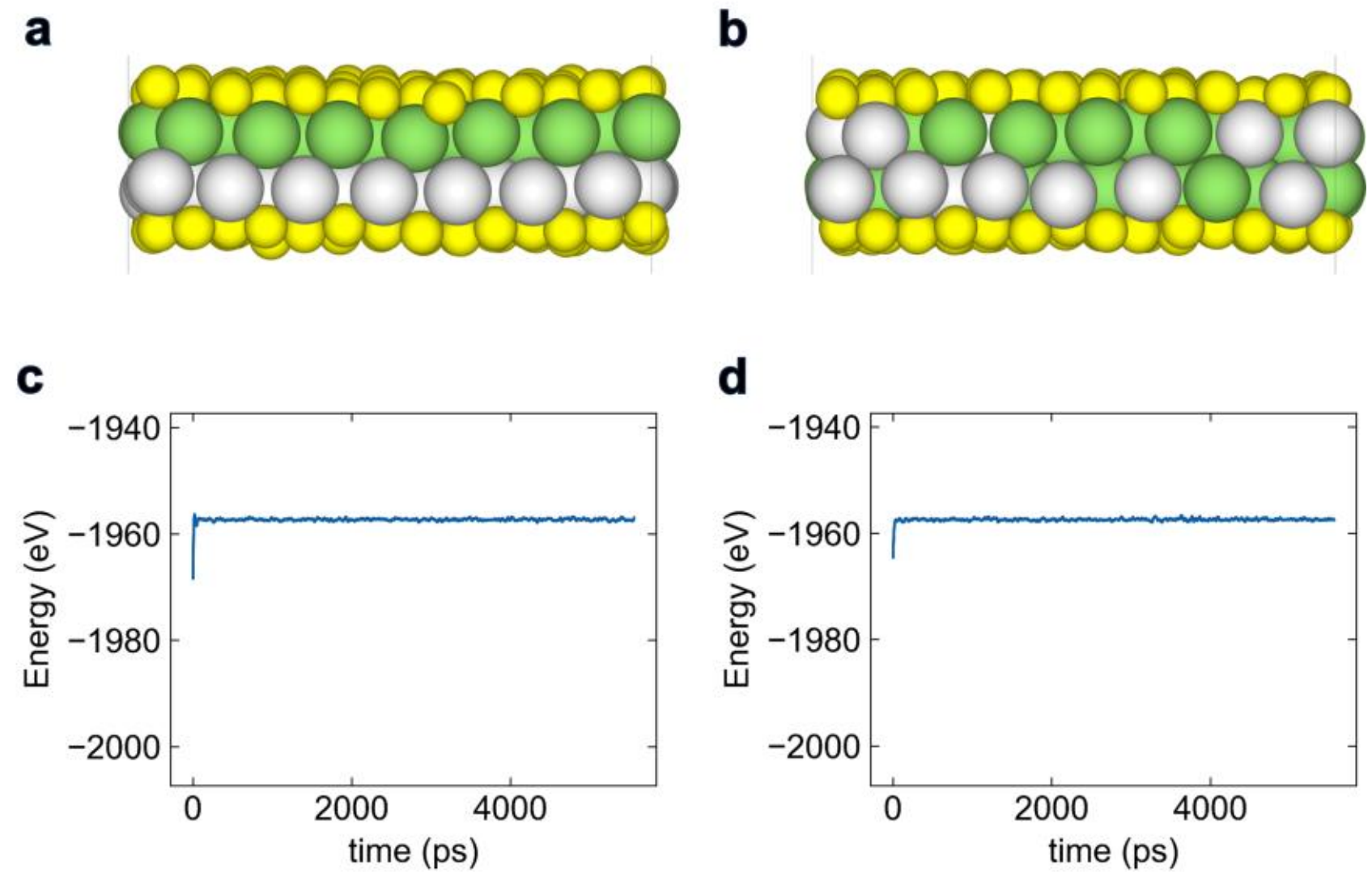

**FIG. S9:** Stability analysis of SMMS intermediate structures. (a-b) Atomic configurations of (a) SMoWS, and (b) alloyed $S(Mo_{0.5}W_{0.5})(W_{0.5}Mo_{0.5})S$ structures. (c-d) Corresponding energy evolution during 5500 ps MLP-MD simulations at 1100 K for (c) SMoWS, and (d) alloyed $S(Mo_{0.5}W_{0.5})(W_{0.5}Mo_{0.5})S$ structures, demonstrating their long-term stability. Note: The data has been smoothed using a low-pass filter for clarity.

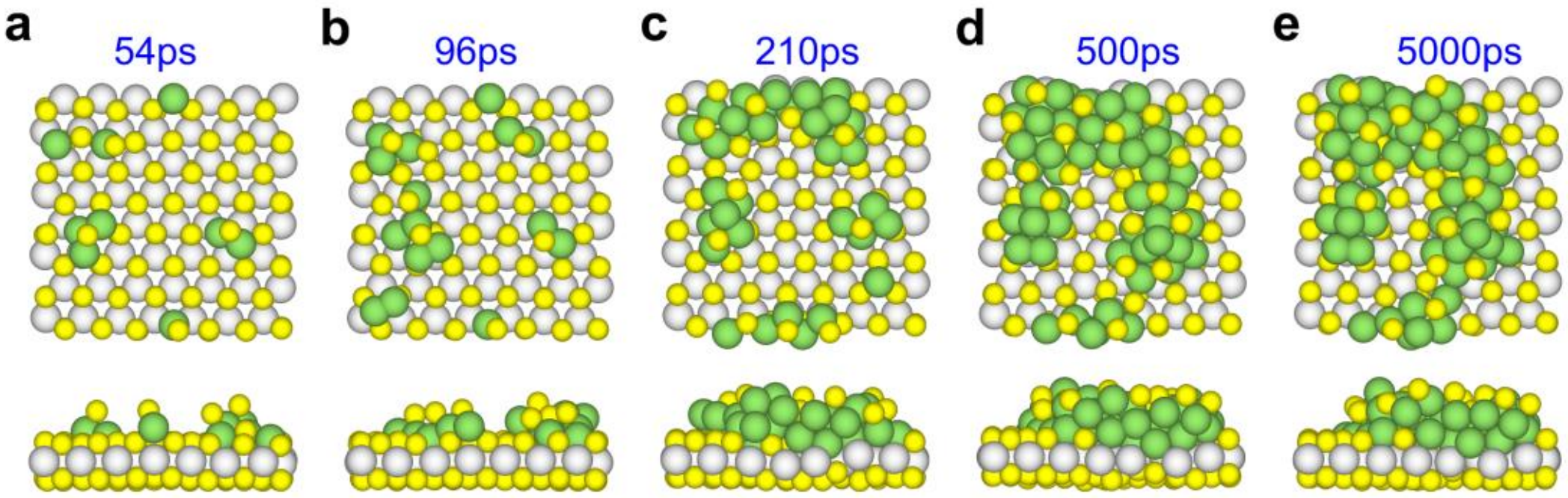

**FIG. S10:** (a-e) Snapshots of the growth structure of $MoS_2/WS_2$ vdWHs during the two-step vapor-deposition process (Mo atoms are deposited on $WS_2$) at 300 K.

## 4. SMMS Intermediate Structures to Bilayer TMDs

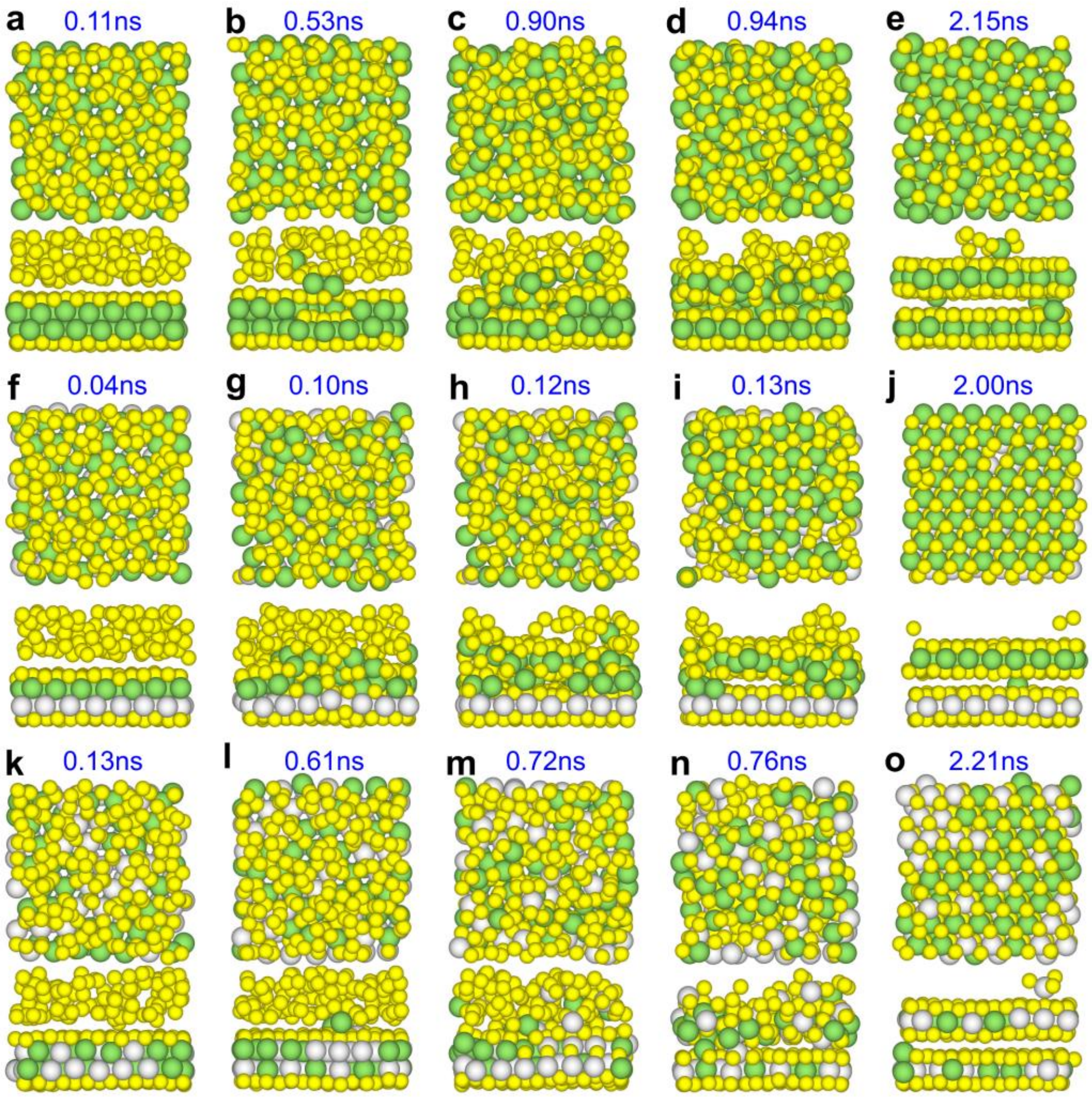

**FIG. S11:** (a-e) MLP-MD simulation with a large number of S atoms placed on top of the SMoMoS structure. The S atoms pull out Mo atoms, forming a bilayer $MoS_2$ structure (1100 K). (f-j) MLP-MD simulation with a large number of S atoms placed on top of the alloyed SMoWS structure. The S atoms pull out Mo atoms, forming a $MoS_2/WS_2$ vdWHs structure (1100 K). (k-o) MLP-MD simulation with a large number of S atoms placed on top of the alloyed $S(Mo_{0.5}W_{0.5})(W_{0.5}Mo_{0.5})S$ structure. The S atoms pull out Mo/W atoms, forming a bilayer alloyed $Mo_{0.5}W_{0.5}S_2$ structure (1100 K).

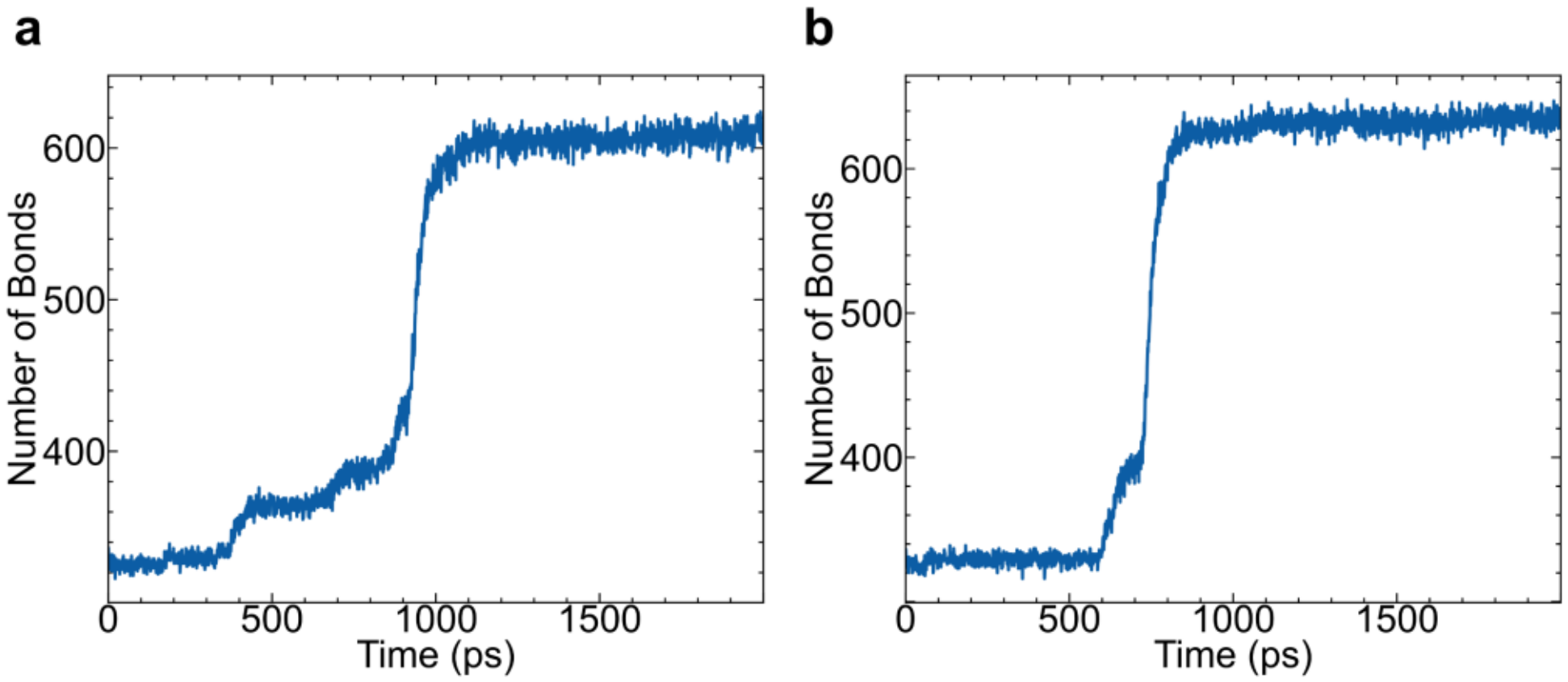

**FIG. S12:** (a) Evolution of the number of Mo-S bonds during sulfur deposition on the SMoMoS intermediate structure. Excess sulfur atoms extract Mo atoms from the SMoMoS structure, leading to the formation of a bilayer $MoS_2$ heterostructure. (b) Evolution of the number of Mo/W-S bonds during sulfur deposition on the alloyed $S(Mo_{0.5}W_{0.5})(W_{0.5}Mo_{0.5})S$ structure. Sulfur atoms facilitate simultaneous extraction of both Mo and W atoms, ultimately forming an alloyed bilayer $Mo_{0.5}W_{0.5}S_2$ configuration. A cutoff distance of 2.6 Å was applied for both Mo-S and W-S bond identifications. The bond counts include both pre-existing and newly formed bonds in the system. The initial number of bonds reflects the presence of metal–sulfur coordination in the intermediate structures, and as the bilayer TMD forms, the number of Mo–S bonds nearly double—indicating the completion of coordination environments in the emerging bilayer structure.

## 5. Simulation of MOCVD TMDs Growth

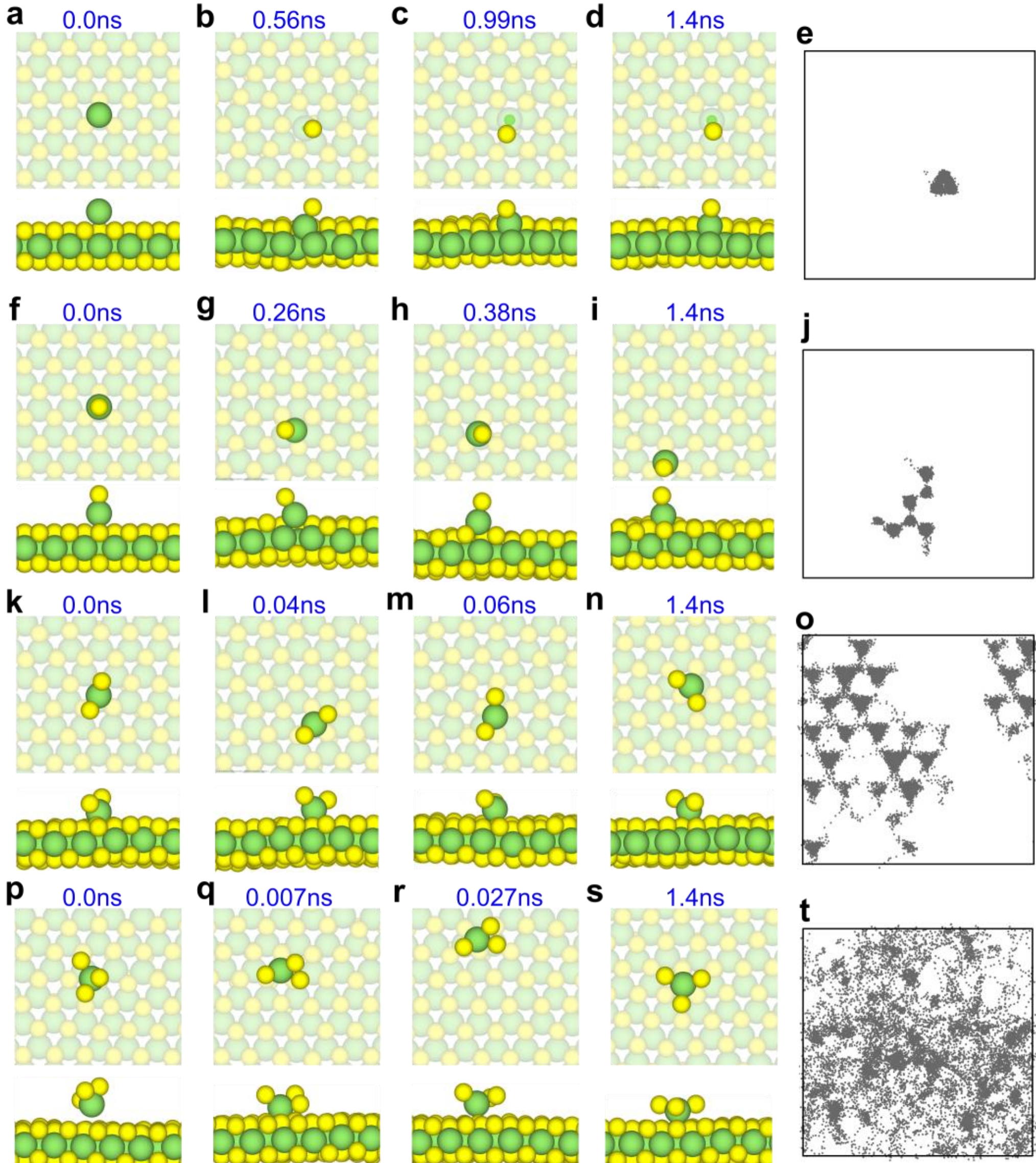

**FIG. S13:** MLP-MD simulations of various Mo-based clusters on $MoS_2$ (1100 K): (a-d) Single Mo atom deposition; (e) Mo atom trajectory on the ***xy*** plane; (f-j) Mo-S cluster; (k-o) Mo-$S_2$ cluster; (p-t) Mo-$S_3$ cluster.

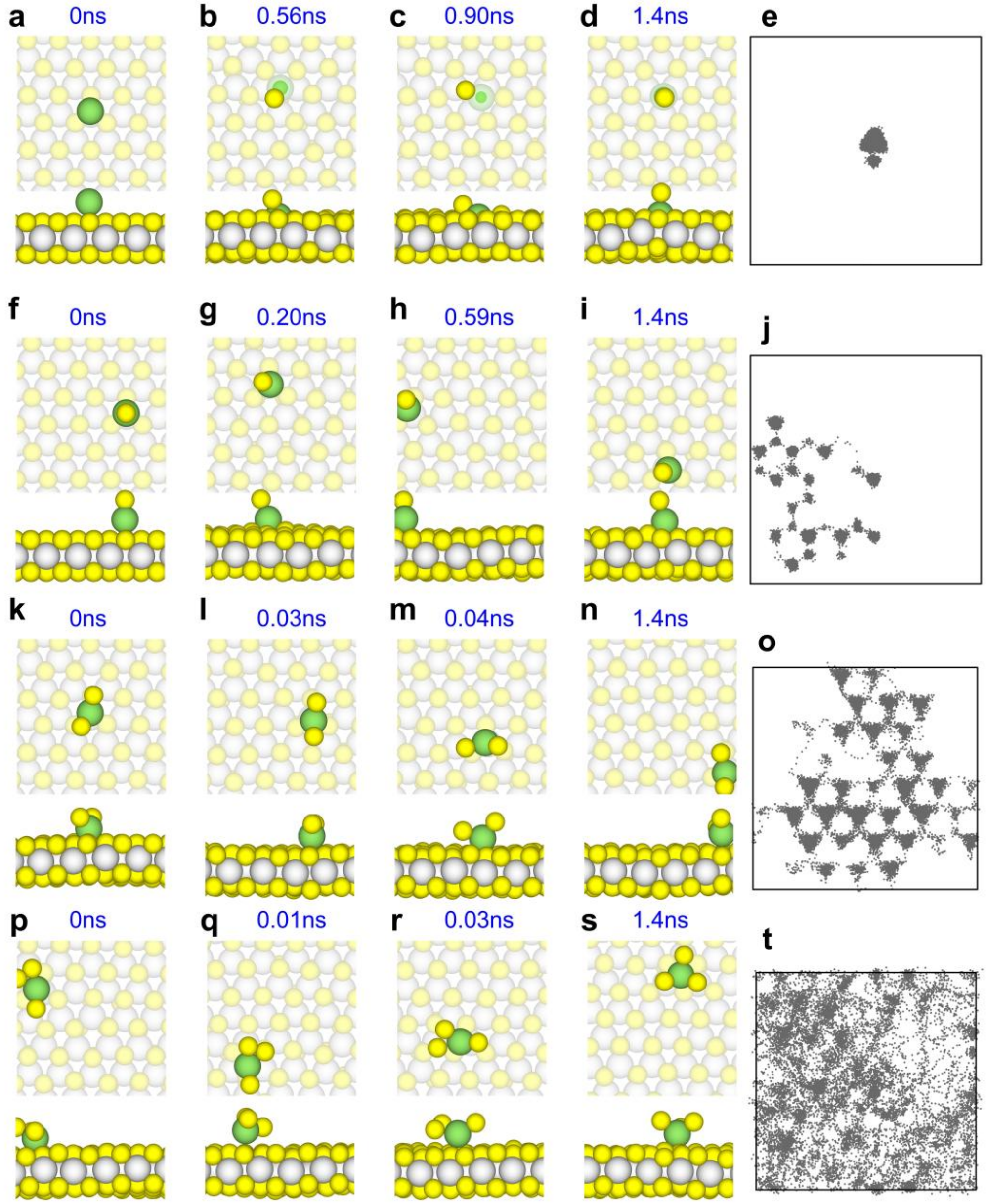

**FIG. S14:** MLP-MD simulations of various Mo-based clusters on $WS_2$ (1100 K): (a-d) Single Mo atom deposition; (e) Mo atom trajectory on the ***xy*** plane; (f-j) Mo-S cluster; (k-o) Mo-$S_2$ cluster; (p-t) Mo-$S_3$ cluster.

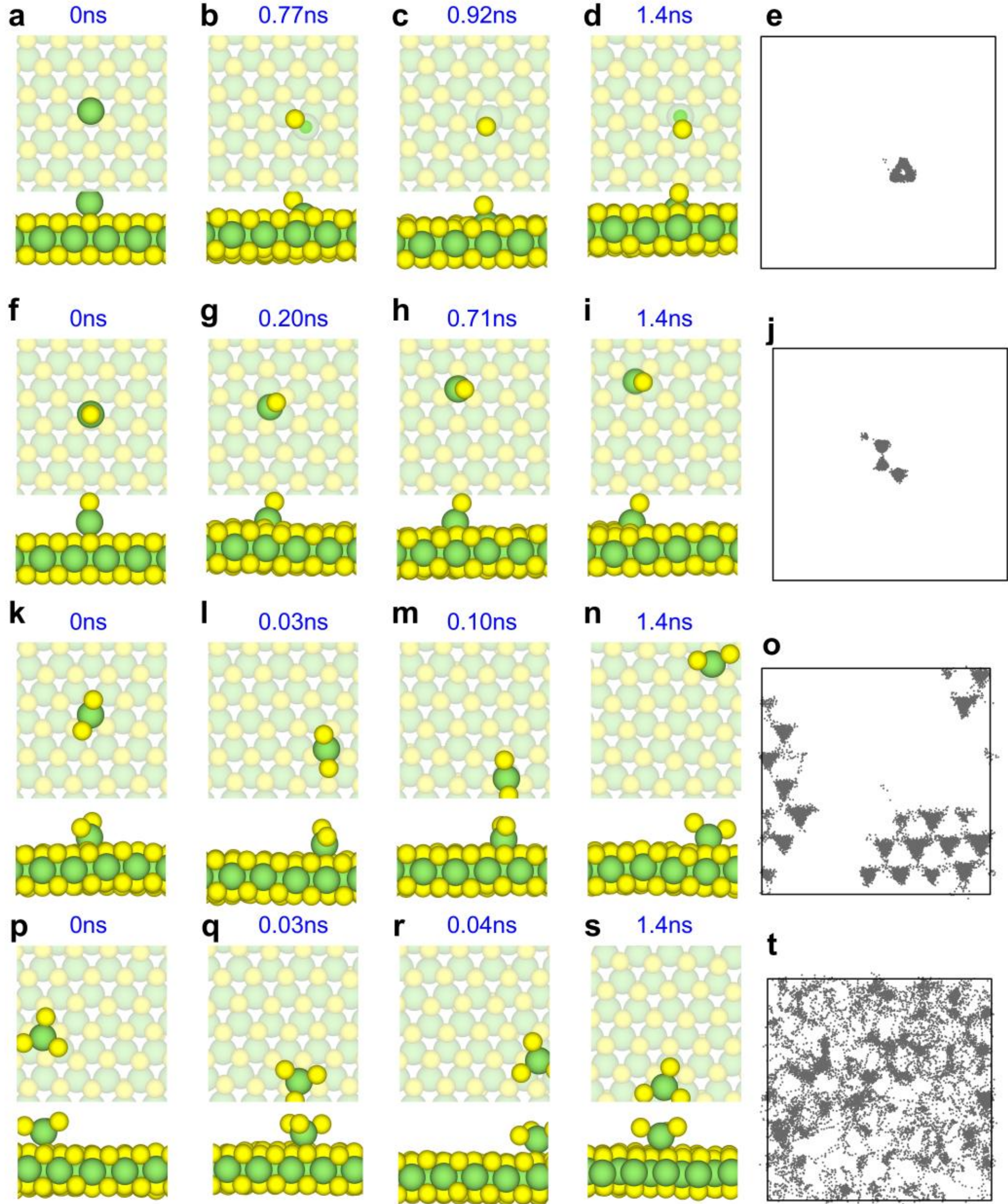

**FIG. S15:** MLP-MD simulations of various Mo-based clusters on $MoS_2$ (900 K): (a-d) Single Mo atom deposition; (e) Mo atom trajectory on the ***xy*** plane; (f-j) Mo-S cluster; (k-o) Mo-$S_2$ cluster; (p-t) Mo-$S_3$ cluster.

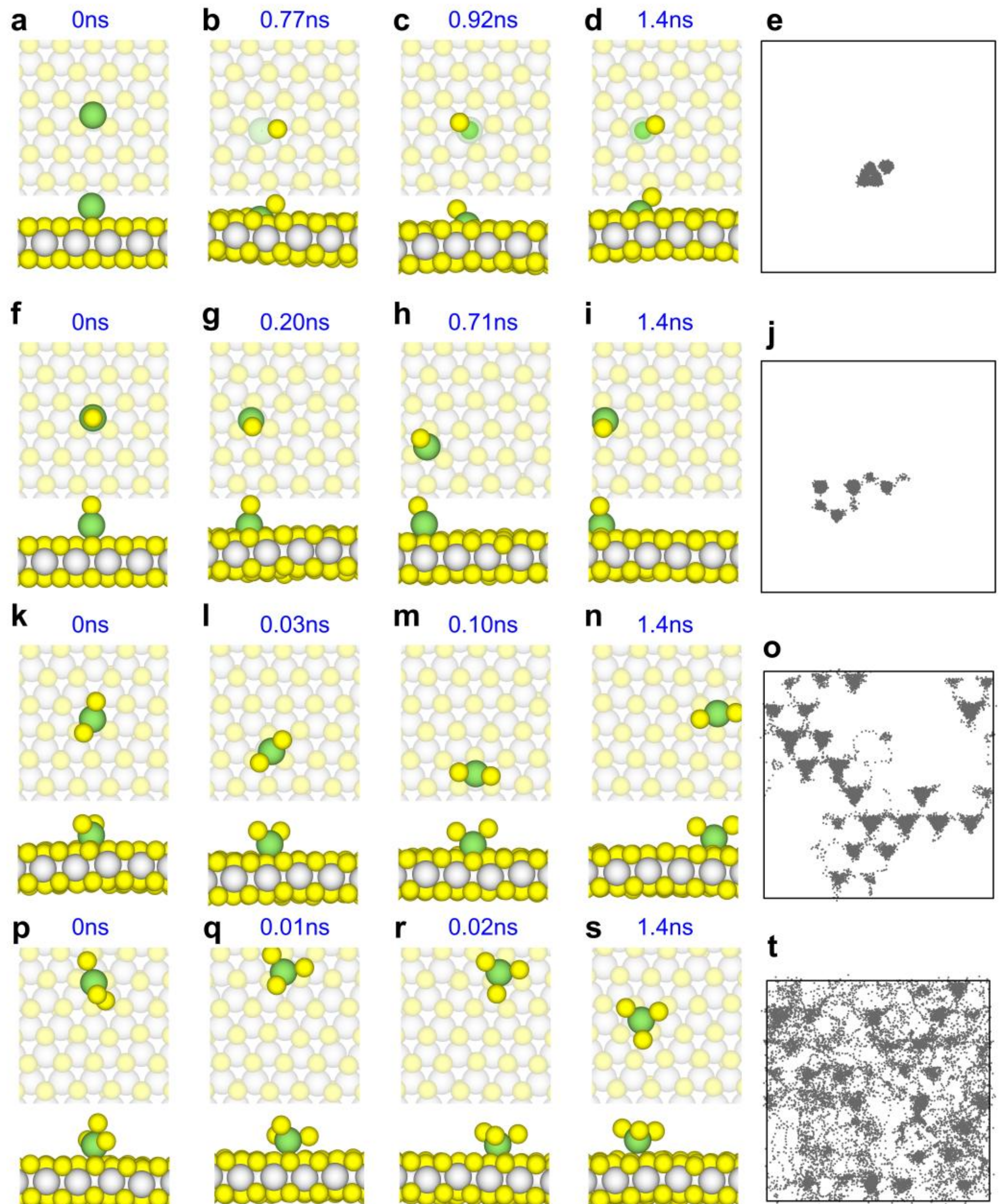

**FIG. S16:** MLP-MD simulations of various Mo-based clusters on $WS_2$ (900 K): (a-d) Single Mo atom deposition; (e) Mo atom trajectory on the ***xy*** plane; (f-j) Mo-S cluster; (k-o) Mo-$S_2$ cluster; (p-t) Mo-$S_3$ cluster.

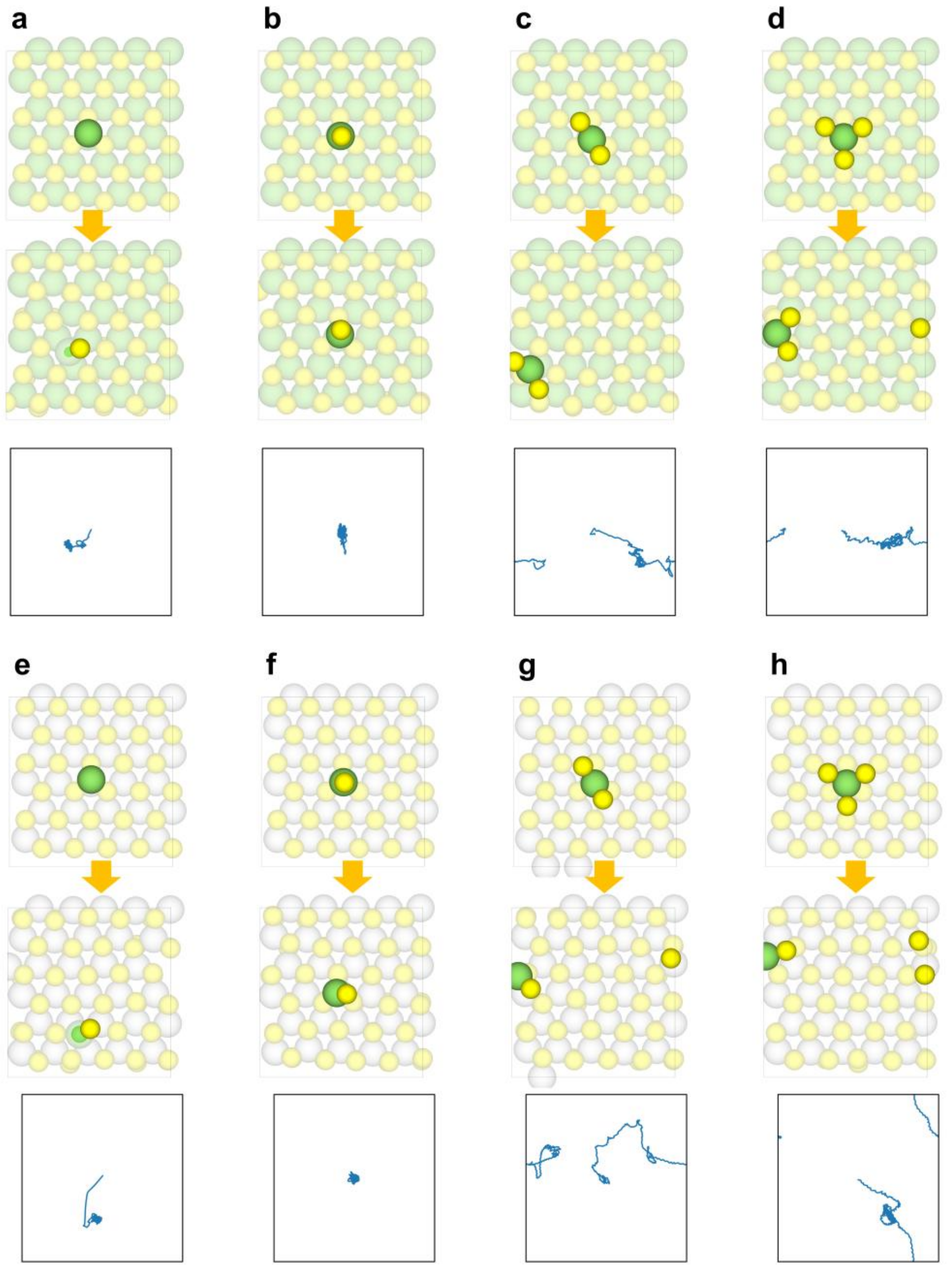

**FIG. S17:** Dynamical behaviour and migration trajectories of Mo-based clusters on monolayer $MoS_2$ and $WS_2$ surfaces (AIMD, 1100 K, 10 ps). Panels (a-d) show the structural evolution of Mo clusters on a $MoS_2$ surface, while (e-h) display the corresponding behaviour on a $WS_2$ surface. Each set of images includes the initial configuration, final configuration, and the in-plane (***xy***) trajectory of Mo atoms. All simulations were performed using AIMD based on the PBE functional under the NVT ensemble at 1100 K for a total duration of 10 ps.

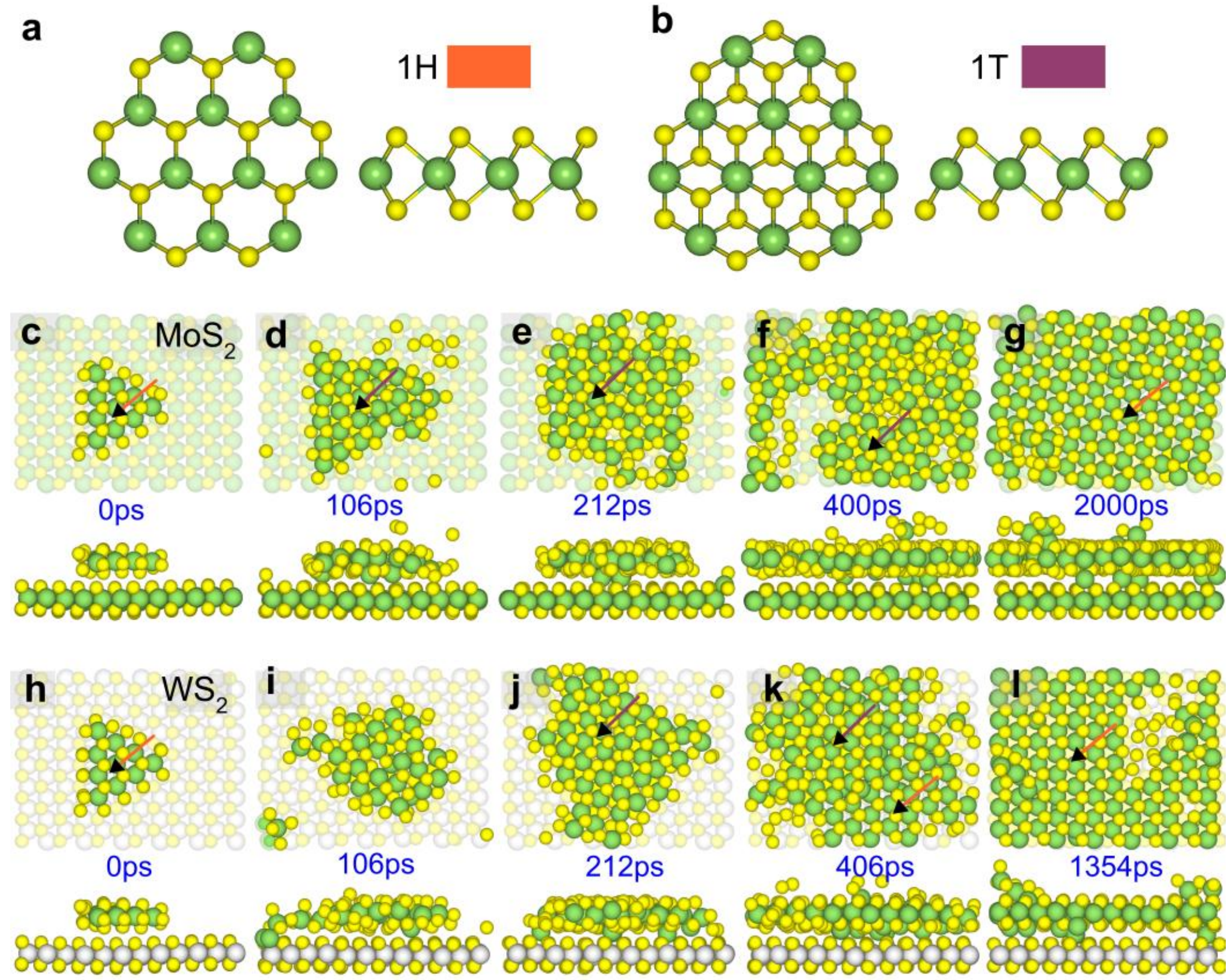

**FIG. S18:** (a-b) Schematic diagrams of (a) 1H-$MoS_2$ and (b) 1T-$MoS_2$ structures. MD simulations illustrating the growth processes of the second $MoS_2$ layer on different substrates: (c-g) $MoS_2$ monolayer substrate and (h-l) $WS_2$ monolayer substrate. The newly formed 1H-phase and 1T-phase domains during growth are indicated by orange and purple arrows, respectively (1100K).

## 6. Electronic Properties of the SMMS Contact with $MoS_2$

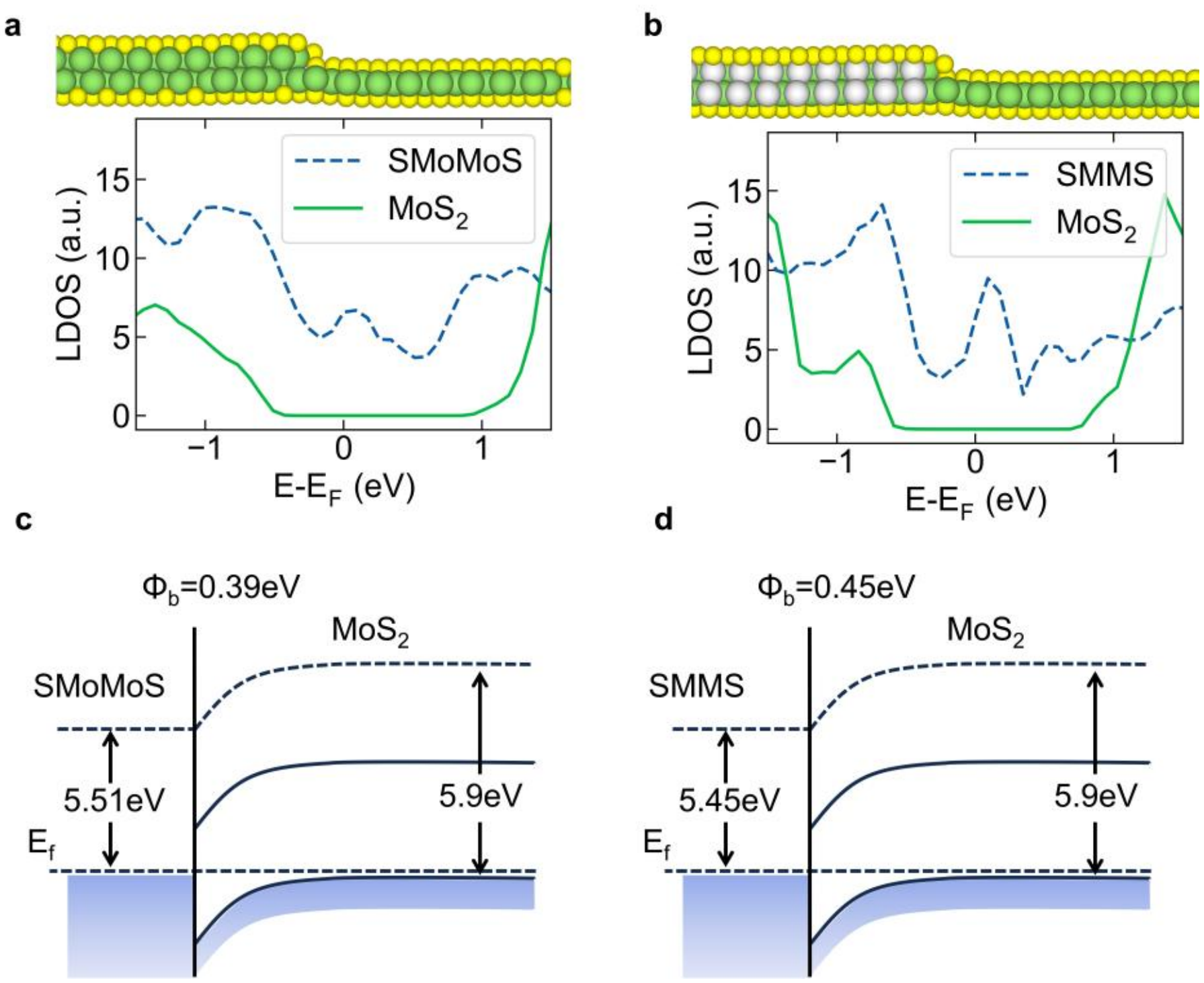

**FIG. S19:** Interfacial electronic properties of TMD heterostructures calculated using PBE functional. LDOS analysis at the interfaces: (a) SMoMoS-$MoS_2$ interface (SBH≈0.5 eV) and (b) alloyed SMMS-$MoS_2$ interface (SBH≈0.6 eV). Upper panels display atomic configurations (S: yellow, Mo: green, W: white). Lower panels show LDOS comparison between intermediate structures (SMoMoS/SMMS: blue dashed lines) and $MoS_2$ (green solid lines), with Fermi level aligned at 0 eV. Schottky-Mott limit analysis for p-type contacts: (c) SMoMoS-$MoS_2$ interface exhibits 0.39 eV SBH, (d) SMMS-$MoS_2$ interface exhibits 0.45 eV SBH.

To investigate the electronic properties and potential applications of the identified intermediate structures, we constructed heterostructures and calculated the p-type Schottky barrier height (SBH) using the PBE functional [2].

FIG. S19(a) shows the LDOS of the SMoMoS-$MoS_2$ interface, while FIG. S19(b) displays the LDOS of the alloyed SMMS-$MoS_2$ interface. The top panels illustrate the atomic configurations of these interfaces, with yellow, green, and white spheres representing S, Mo, and W atoms, respectively. In the bottom panels, the calculated LDOS is presented, where blue dashed lines represent the LDOS of the intermediate structures (SMoMoS or SMMS), and green solid lines represent the LDOS of $MoS_2$. The Fermi level is set to 0 eV.

Analysis of these LDOS profiles reveals the formation of p-type Schottky barriers at both interfaces. For the SMoMoS-$MoS_2$ interface (FIG. S19(a)), we estimate a SBH of approximately 0.5 eV. In contrast, the alloyed SMMS-$MoS_2$ interface (FIG. S19(b)) exhibits a slightly higher barrier of about 0.6 eV. These findings are significant as they suggest that these intermediate structures, particularly SMoMoS, could potentially serve as low-resistance contacts for $MoS_2$-based electronic devices. The relatively low SBHs indicate favourable charge transport characteristics, which could be advantageous for various applications in nanoelectronics and optoelectronics utilizing TMD heterostructures.